\documentclass[prd,byrevtex
,preprint
,amsfonts,amsmath,amssymb
]{revtex4}
\usepackage[T1]{fontenc}
\usepackage{comment}
\usepackage[dvipdfmx]{graphicx}
\usepackage{float}
\usepackage{color}
\usepackage{amsmath,amsthm,amssymb}
\usepackage{bm}
\usepackage[dvipdfmx]{graphicx}
\usepackage{ascmac}
\usepackage{braket}
\usepackage{latexsym}
\usepackage{cases}
\newcommand{\hs}{\hspace{-.1em}}
\newcommand{\ti}{\tilde}
\newcommand{\beq}{\begin{equation}}
\newcommand{\eeq}{\end{equation}}
\newcommand{\bea}{\begin{eqnarray}}
\newcommand{\eea}{\end{eqnarray}}
\newcommand{\Tr}{{\rm Tr}}
\newcommand{\dg}{\dagger}
\newcommand{\nnm}{\nonumber}
\newcommand{\Bl}{\Bigl}
\newcommand{\Br}{\Bigr}
\newcommand{\bl}{\bigl}
\newcommand{\br}{\bigr}
\newcommand{\bmx}{\left[\begin{array}}
\newcommand{\emx}{\end{array}\right]}
\newcommand{\bfg}{\begin{figure}}
\newcommand{\efg}{\end{figure}}
\newcommand{\bmn}{\begin{minipage}}
\newcommand{\emn}{\end{minipage}}
\newcommand{\bnmc}{\begin{numcases}}
\newcommand{\enmc}{\end{numcases}}
\newcommand{\bca}{\begin{cases}}
\newcommand{\eca}{\end{cases}}
\newcommand{\bfr}{\begin{flushright}}
\newcommand{\efr}{\end{flushright}}
\newcommand{\bfl}{\begin{flushleft}}
\newcommand{\efl}{\end{flushleft}}
\newcommand{\bct}{\begin{center}}
\newcommand{\ect}{\end{center}}
\newcommand{\incg}{\includegraphics}
\newcommand{\pa}{\partial}
\newcommand{\mcl}{\mathcal}
\newcommand{\dip}{\displaystyle}

\newcommand{\Dlt}{\Delta}
\newcommand{\dlt}{\delta}

\newcommand{\tht}{\theta}
\newcommand{\fr}{\frac}
\newcommand{\sq}{\sqrt}
\newcommand{\eps}{\epsilon}
\newcommand{\omg}{\omega}

\newcommand{\bkt}{\braket}

\newcommand{\h}{\hat}
\newcommand{\al}{\alpha}

\begin{document}
\title{Parameter estimation by decoherence in the double-slit experiment }
\author{Akira Matsumura}\email{matsumura.akira@h.mbox.nagoya-u.ac.jp}
\author{Taishi Ikeda}
\author{Shingo Kukita}
\affiliation{Department of Physics, Graduate School of Science, Nagoya 
University, Chikusa, Nagoya 464-8602, Japan}
\begin{abstract}
We discuss a parameter estimation problem using quantum decoherece in the double-slit interferometer. We consider a particle coupled to a massive scalar field after the particle passing through the double slit and solve the dynamics non-perturbatively for the coupling by the WKB approximation. This allows us to analyze the estimation problem which cannot be treated by master equation used in the research of quantum probe. In this model, the scalar field reduces the interference fringes of the particle and the fringe pattern depends on the field mass and coupling. To evaluate the contrast and the estimation precision obtained from the pattern, we introduce the interferometric visibility and the Fisher information matrix of the field mass and coupling. For the fringe pattern observed on the distant screen, we derive a simple relation between the visibility and the Fisher matrix. Also, focusing on the estimation precision of the mass, we find that the Fisher information characterizes the wave-particle duality in the double-slit interferometer.
\end{abstract}
%
\maketitle
\section{INTRODUCTION}
The double-slit experiment is a traditional experiment\cite{1,2,3,4} to observe an interference effect of a quantum particle. In the simplest set-up, particles go through a double slit on a plate, and a screen behind the plate detects the particles. As is well known, the existence of quantum superposition is demonstrated through this experiment; if we individually send particles through the double slit to the detection screen, then the image on the screen has an interference pattern. When an observer detects which slit each particle goes on, then the interference pattern disappears. This is because the wave function of the single particle is collapsed by measuring the path of the particle. The phenomenon is called quantum decoherence, which is important to understand how classical nature emerges from a quantum development\cite{5,6,7}. The quantum decoherence of the focused system is often induced when the system interacts with a large environment system. The experiments and theoretical analysis of quantum decoherence of fullerenes were done by using a Talbot-Lau  interferometer\cite{8,9}. In Ref.\cite{8} and \cite{9}, the quantum decoherence was investigated when fullerenes proceed in a gas (interaction with gas) and emit thermal radiation (couple to the electromagnetic field), respectively.     

Quantum decoherence is important for not only theoretical or conceptual problems, but also practical aspects. In the field if quantum metrology, there are several researches of quantum probe using quantum decoherence has been discussed \cite{10,11,12,13,14,15,16,17,18}. In Ref. \cite{14} and \cite{15}, the authors considered a parameter estimation of an environment system by using a spin system as a probe and evaluated the quantum Fisher information (QFI) about the parameter. ( QFI gives a minimum estimation error for all quantum measurements, which is called the quantum Cramer-Rao bound\cite{19,20,21}.)  Also, in Ref. \cite{18} a single harmonic oscillator as a probe was considered for the parameter estimation of the environment composed of harmonic oscillators. Recently, the parameter estimation of a continuous field has been discussed. In Ref. \cite{23,24}, the estimation by phonons of the amplitude of gravitational waves was studied, and in Ref. \cite{25} the quantum metrology of the Unruh temperature by utilizing two detectors was considered. In this paper, we want to consider the estimation problem of field parameters by quantum decoherence with a different approach not used in previous works.    

In Ref. \cite{14,15,18} mentioned above, the evolution of the probe system was based on the quantum master equation. Hence the analysis was perturbative for the coupling to the environment and has neglected the reaction to the environment. When the coupling constant is large, the environment changes significantly and the quantum master equation cannot be applied to solve the dynamics of the probe system. In general, it is difficult to solve the dynamics of a quantum system interacting with a many body system. However, in Ref.\cite{22} the time evolution of an electron coupled to the vacuum of an electromagnetic field was derived non-perturbatively for the coupling in the WKB approximation. The WKB approximation is applied for the case where the electron is sufficiently localized than the fluctuation scale of the electromagnetic field. In this approximation, the path of the particle is unchanged, however the field fluctuation induces the phase shift and the decoherence to the interference pattern of the electron. By using the analysis, we can discuss the parameter estimation of an environment via the quantum decoherence of the particle, and this strategy can applied for the higher order of the coupling. 

To evaluate the parameter estimation by quantum decoherence, we need to consider a superposed particle coupled to an environment. Hence, we suppose that a particle interacts with an environment after the particle passing through a double slit. For example, this situation corresponds to the experiment setting in Ref. \cite{8,9} or the Casimir effect considered in Ref. \cite{22}. Then we treat the dynamics of the particle non-perturbatively for the coupling and discuss a estimation problem of the environment parameters using the interference pattern obtained in the double-slit experiment. To get some analytical results and understand the behavior of the quantum decoherence, we consider a one-dimensional non-relativistic particle (as a probe) coupled to a massive scalar field (as an environment) for a finite time. The initial state of the particle is the superposition of two Gaussian wave packets and the time evolution is calculated. The probability distribution of the particle corresponds to the histogram of the incident particle observed on the screen, and the interference pattern depends on the field mass and coupling. We evaluate the interferometric visibility\cite{25} which quantifies the contrast of interference and the Fisher information (FI) matrix of the mass and coupling from the probability distribution. The FI matrix gives the lower bound of the estimation precision for position measurement. Through this analysis, we derive a simple relation between the visibility and FI matrix. The simple relation tells us how the quantum decoherence in the double-slit experiment depends on the parameters of the environment. Furthermore, by focusing on FI of the field mass, we find that FI depending on the quantum decoherence in the double-slit interferometer characterizes the wave-particle duality.

This paper is organized as follows: in Sec.I\hs{I} we treat a toy model of the double-slit experiment and calculate the probability distribution of a non-relativistic particle. This probability distribution describes the histogram of the particle on the screen. In Sec.I\hs{I}\hs{I} we introduce a massive scalar field interacting with the particle for a finite time and approximate its dynamics assuming the interaction time is short. We compute the free evolution of the particle which have interacted for the finite time. The probability density of the particle depends on the field mass and coupling. In Sec.I\hs{V} the interferometric visibility and FI matrix of the mass and coupling are calculated from the probability density. Then we derive a relation between the visibility and FI matrix under some approximations. In Sec.V we discuss the behaviors of the visibility and FI of each parameter. Sec.V\hs{I} is devoted to conclusion. In this paper, we use the natural units; $\hbar=c=1$.

\section{Toy model of the double-slit experiment}
In this analysis, a toy model of the double-slit experiment is described as the one-dimensional system of a non-relativistic free particle. In App.A, we show that the double-slit experiment is analyzed as the one-dimensional system of a non-relativistic particle by choosing a class of initial conditions. 

The Hamiltonian of the non-relativistic free particle is 
\beq
\h{H}_{\rm{p}}=\fr{\h{p}^{2}}{2m_{\rm{p}}}, 
\eeq
where $\h{p}$ is the momentum of the particle and $m_{\rm{p}}$ is the particle mass.  We use the initial state of the particle
\beq
\ket{\Psi_{\rm{in}}}=A\bl(\ket{\psi_{+}}+\ket{\psi_{-}}\br),\psi_{\pm}(x)(=\bkt{x|\psi_{\pm}})=\fr{1}{(\pi \Dlt^{2})^{1/4}}\exp \Bl[-\fr{(x\mp L)^{2}}{2\Dlt^{2}} \Br],
\eeq
to identify the state of a particle immediately after passing through the slits ($x=\pm L$ in Fig.\ref{fig1}). $A$ is the normalization given by $(2+2e^{-L^{2}/\Dlt^{2}})^{-1/2}$ and $\Dlt$ corresponds to the size of each slit. In the following, all variables and parameters are normalized by $\Dlt$.
\bfg[htbp]
    \incg[clip,height=7cm,width=11.0cm]{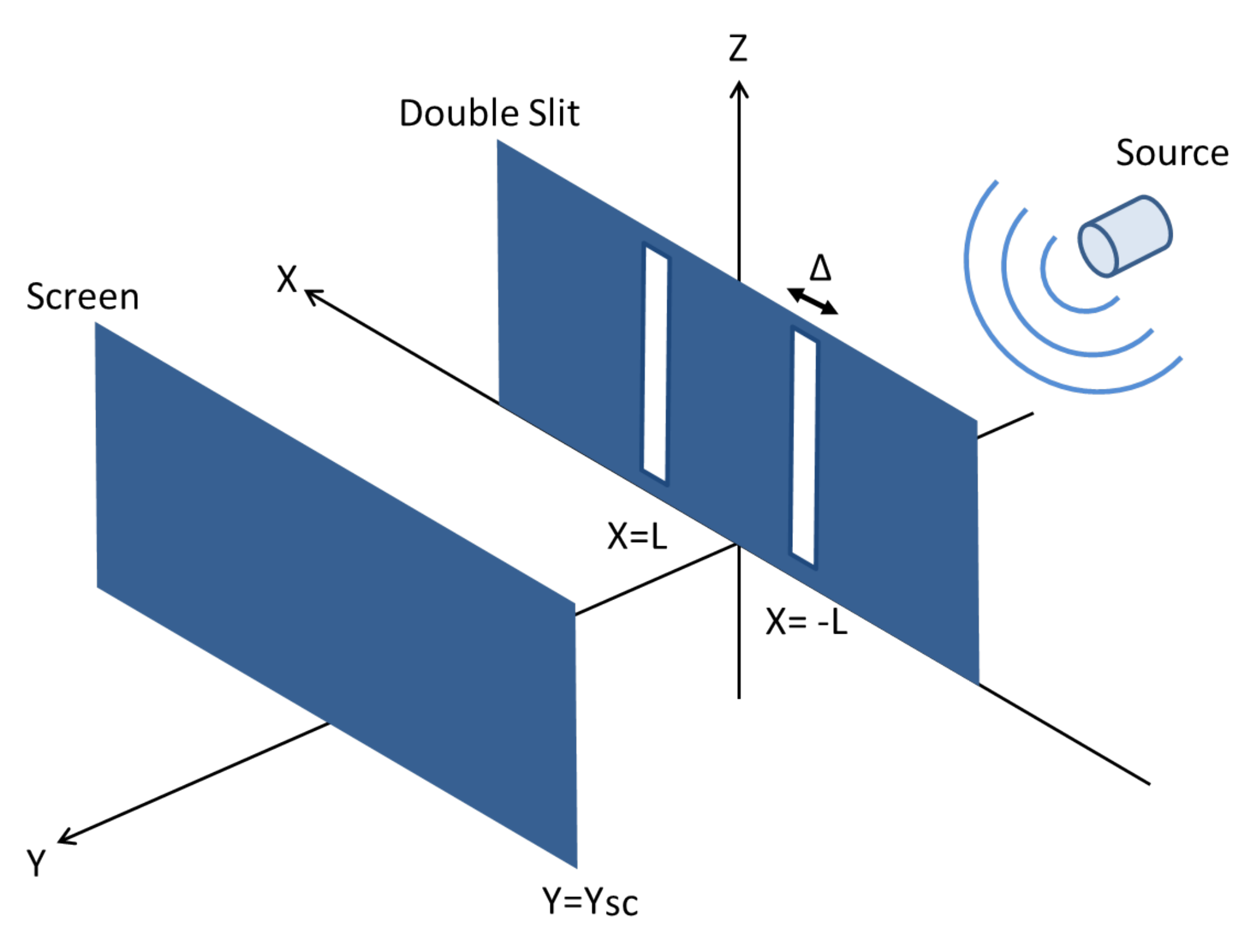}
    \caption{The setup of the double-slit experiment, which has a source and two plates. Two slits ($x=\pm L$) are on the first plate in front of the source. The size of each slit is $\Dlt$. The second plate at $y=y_{\rm{sc}}$ is a detection screen for the particle which have passed through the slits.}
    \label{fig1}
\efg
After the free evolution, the probability density for $x$ at time $t$ is given by 
\bea
p_{0}(x,t)&=&|\bra{x}e^{-i\h{H}_{\rm{p}}t}\ket{\Psi_{\rm{in}}}|^{2} \nnm \\
&=&A^{2}\Bl(|\psi_{+}(x,t)|^{2}+|\psi_{-}(x,t)|^{2} +\psi_{+}(x,t)\psi^{*}_{-}(x,t)+\psi^{*}_{+}(x,t)\psi_{-}(x,t)\Br),
\eea
where $\psi_{\pm}(x,t)$ is given by
\bea
&&\psi_{\pm}(x,t)
=\int dx' G(x-x',t) \psi_{\pm}(x') =\fr{1}{\pi ^{1/4}\sq{1+i t/m_{\rm{p}}}} \exp \Bl[-\fr{(x\mp L)^{2}}{2(1+i t/m_{\rm{p}})} \Br],\\
&&G(x-x',t-t')
=\sq{\fr{m_{\rm{p}}}{2\pi i (t-t')}}\exp \Bl[-i\fr{m_{\rm{p}}(x-x')^{2}}{2(t-t')} \Br].
\eea
The probability density $p_{0}(x,t)$ represents the histogram of the particles on the screen at an observation time $t$ (See App. A). Fig.\ref{fig2} shows $p_{0}/A^{2}$ as a function of $x$ for $L=10$ and $ t/m_{\rm{p}}=0,20$. In the right panel of FIG. \ref{fig2}, we observe that the function $p_{0}/A^{2}$ for $ t/m_{\rm{p}}=20$ oscillates for $x$. This behavior corresponds to the fringe pattern on the screen in the double-slit experiment. In the next section, we consider a massive scalar interacting with the particle, and calculate the probability distribution of the particle.
\bfg[htbp]
    \incg[clip,height=4.5cm,width=14.0cm]{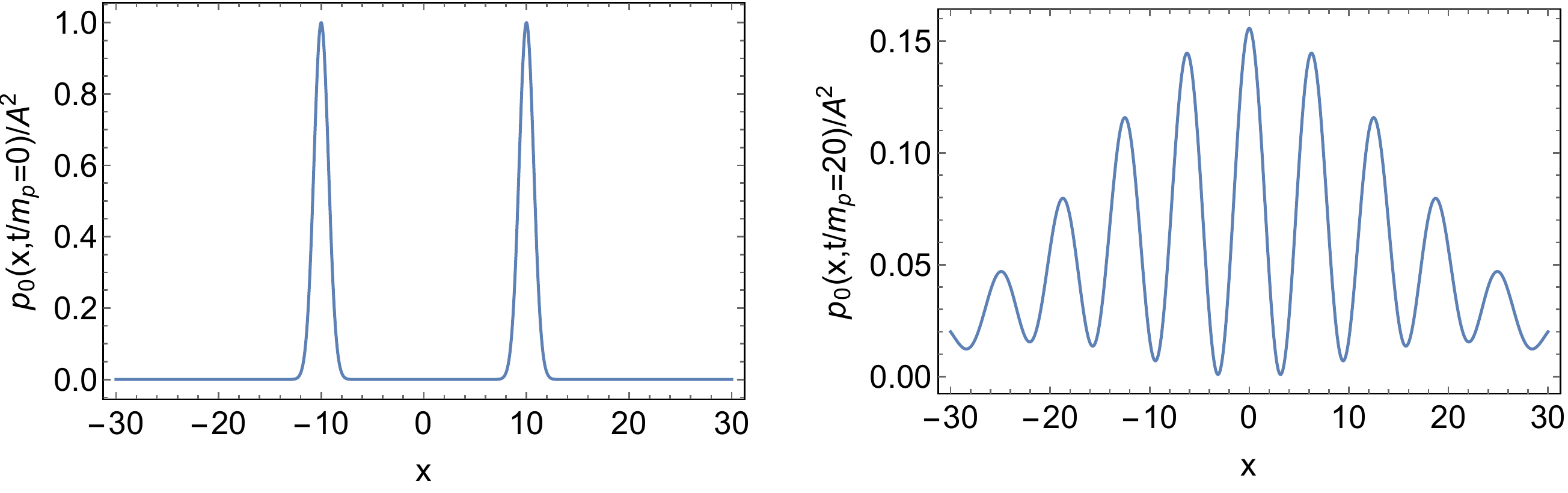}
    \caption{The behavior of $p_{0}/A^{2}$ as a function of $x$ for  $ L=10$ and $t/m_{\rm{p}}=0,20 $. The behavior of the function represents the interference pattern on the screen in the double slit experiment.}
    \label{fig2}
\efg

\section{Toy model of the double-slit experiment : the particle coupled to a massive scalar field}
Let us introduce a massive scalar field interacting with a non-relativistic particle and calculate the time evolution of the particle. This is a just toy model, however the choice of the environment can give some analytical solutions and the typical behavior about quantum decoherence. The free Hamiltonian of the scalar field $\h{\phi}(x)$ and the interaction Hamiltonian are  
\bea
\h{H}_{\phi}&=&\int dx \fr{1}{2}\Bl(\h{\pi}(x)^{2}+(\pa_{x}\h{\phi}(x))^{2}+ m^{2}\h{\phi}^{2}(x) \Br), \\
\h{H}_{\rm{int}}&=&g\int dx \dlt(\h{x}-x) \h{\phi}(x),
\eea
where $\h{\pi}(x)=\pa \h{\phi}/\pa t$ is the conjugate momentum of the field $\h{\phi}(x)$ and $m$ is the mass of the scalar field. $g$ is the coupling constant of the interaction and $\h{x}$ is the position operator of the particle.  however 
The canonical variables $(\h{\phi},\h{\pi})$ follow the canonical commutation relations
\beq
\bl[\h{\phi}(x),~\h{\pi}(x')\br]=i\dlt(x-x'),~others=0.
\eeq
We consider that the system is initially the product state  
\beq
\ket{\Psi_{\rm{in}}}=A \bl(\ket{\psi_{+}}+\ket{\psi_{-}}) \ket{0_{\phi}},
\eeq
where $\ket{0_{\phi}}$ is the ground state of the free Hamiltonian $H_{\phi}$. We assume that the particle interacts with the scalar field for a finite time. The interaction time corresponds to a time during which the particle passes through the inside of a many-body system or a continuous field. In the interaction picture, the time evolution of the state from $0$ to $T$ is given by
\beq
A \bl(\ket{\psi_{+}}+\ket{\psi_{-}}) \ket{0_{\phi}} \longrightarrow{\rm{T_{p}} \exp}\bl[-i\int^{T}_{0} dt \h{H}_{\rm{int}}(t) \br]A \bl(\ket{\psi_{+}}+\ket{\psi{-}}) \ket{0_{\phi}},
\eeq 
where $\rm{T_{p}}$ is the time-ordered product. If the interaction time $T$ is much smaller than the typical time scale of the free non-relativistic particle $m_{\rm{p}}$, then the time evolution is evaluated as 
\beq
{\rm{T_{p}}} \exp\bl[-i\int^{T}_{0} dt H_{\rm{int}}(t) \br]A \bl(\ket{\psi_{+}}+\ket{\psi_{-}}) \ket{0_{\phi}}\approx A\int dx \bl(\psi_{+}(x)+\psi_{-}(x)\br) \ket{x} \h{U}(x,T) \ket{0_{\phi}} \label{13}
\eeq
where $\ket{x}$ is the eigenstate of $\h{x}$ and $\h{U}(x,t)$ is defined by 
\beq
\h{U}(x,t)={{\rm{T_{p}} \exp}}\bl[-ig \int^{t}_{0} dt'\h{\phi}(x,t') \br]. \label{15}
\eeq 
The explicit derivation of the formula (\ref{13}) is given in App. B. After the interaction with the scalar field, the reduced density operator $\h{\rho}^{I}$ (``~I~'' means the interaction picture) of the particle is give as
\beq
\h{\rho}^{I}\approx A^{2}\int^{\infty}_{-\infty} dx'dx \Bl(\psi^{*}_{+}(x')+\psi^{*}_{-}(x') \Br) \Bl(\psi_{+}(x)+\psi_{-}(x) \Br)   \bra{0_{\phi}} \h{U}^{\dg}(x',T)\h{U}(x,T) \ket{0_{\phi}}\ket{x}\bra{x'} \nnm \\.
\eeq
The inner product $\bra{0_{\phi}} \h{U}^{\dg}(x',t)\h{U}(x,t) \ket{0_{\phi}}$ was calculated in \cite{22}. The formula is 
\beq
\bra{0_{\phi}}U^{\dg}(x',t)U(x,t) \ket{0_{\phi}}=\exp \bl[ -W(x-x',t)\br], 
\eeq
where $W$ is given as 
\bea
W(x-x',t)&=&\fr{g^{2}}{2}\int^{t}_{0} dt' \int^{t}_{0} dt'' \Bl( \bra{0_{\phi}}{\rm{T_{p}}}\bl[\h{\phi}(x,t')\h{\phi}(x,t'')\br]^{\dg}\ket{0_{\phi}}  \nonumber \\
&+&\bra{0_{\phi}}{\rm{T_{p}}}\bl[\h{\phi}(x',t')\h{\phi}(x',t'')\br] \ket{0_{\phi}} -2\bra{0_{\phi}}\h{\phi}(x,t')\h{\phi}(x',t'') \ket{0_{\phi}}\Br), \label{14}
\eea
and we explain the meaning of each term in the equation (\ref{14}) in FIG.\ref{fig3}. The first and second term denote the exchange of the scalar field in each path of the particle. The third term represents the exchange of the scalar field between two paths. 
\bfg[htbp]
 \bmn{0.5\hsize}
  \incg[clip,height=6.0cm,width=5.0cm]{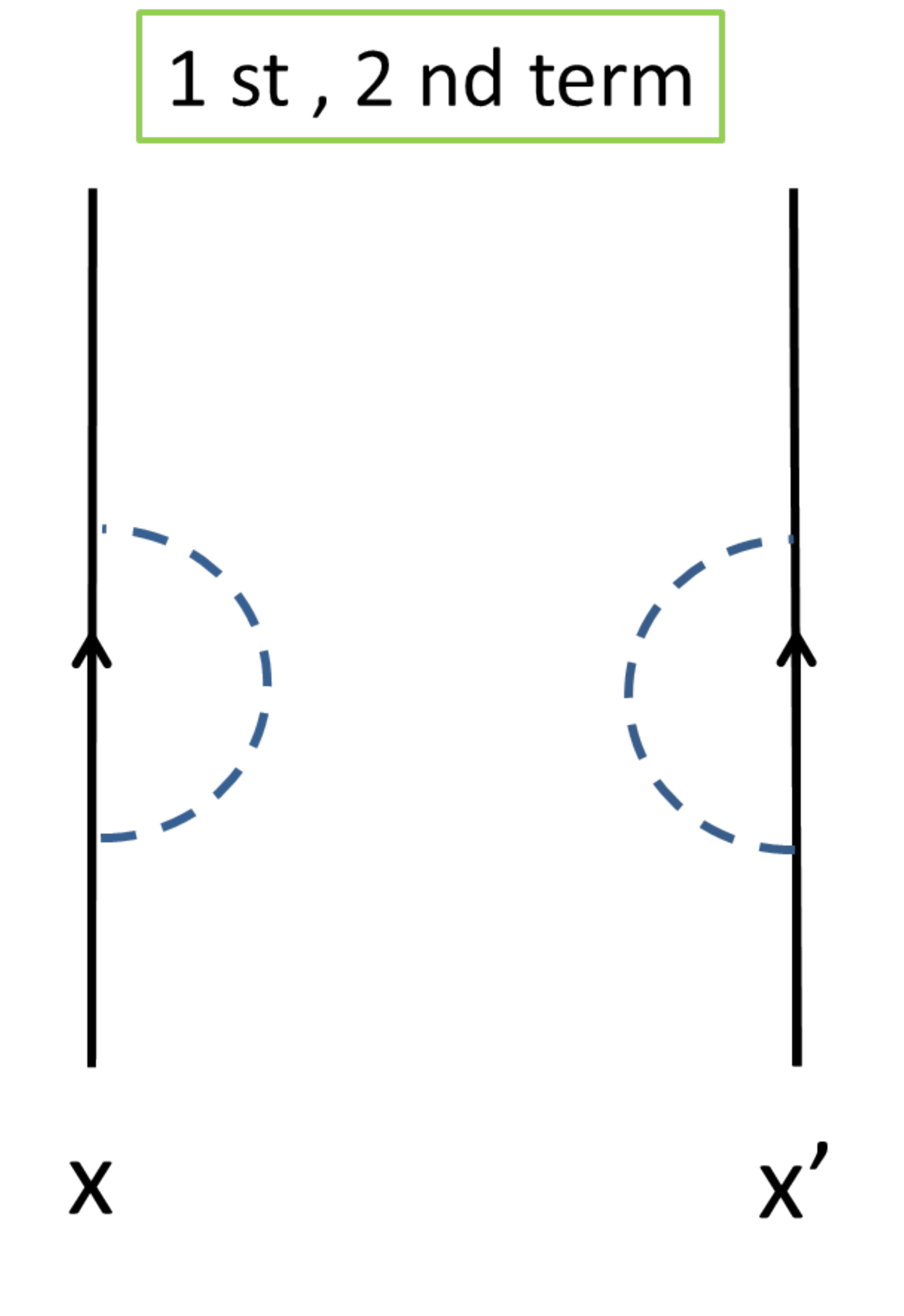}
 \emn
 \bmn{0.5\hsize}
  \incg[clip,height=6.0cm,width=5.0cm]{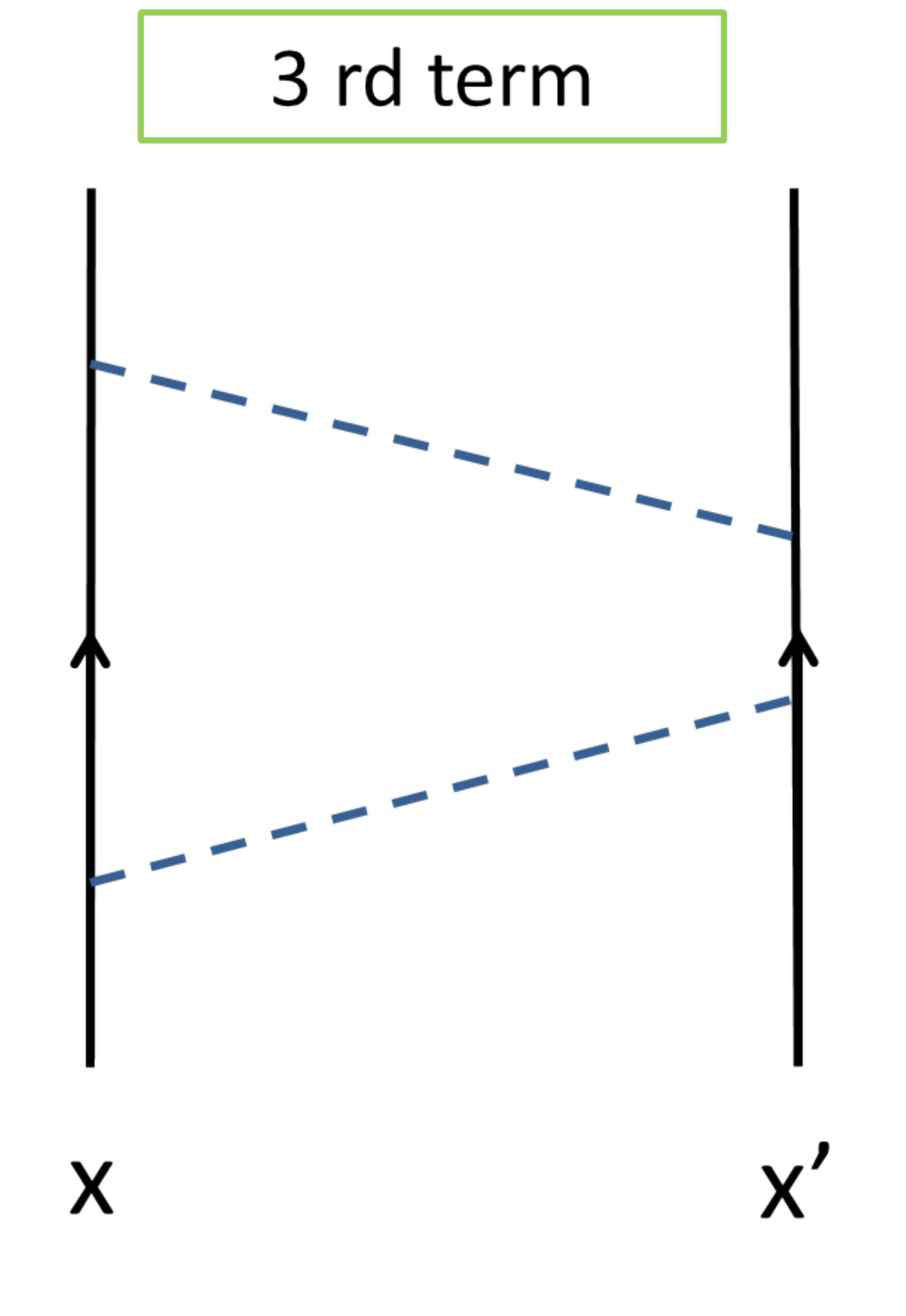}
 \emn
    \caption{The schematic diagrams of the contributions included in the function $W(x-x',t)$. The black line denotes the path of the particle, and the dashed line describes the propagator of the scalar field. The first and second term of the equation (\ref{14}) give the contribution of the exchange of the scalar field in each path. The third term represents the exchange of the scalar field between two paths.}
    \label{fig3}
\efg
In App.C we derive the following formula of the function $W(x-x',t)$:  
\bea
W(x-x',t)&=&\fr{g^{2}}{\pi m^{2}} \int^{\infty}_{0}\fr{du}{(u^{2}+1)^{3/2}}(1-\cos [m(x-x')u])(1-\cos[mt \sq{u^{2}+1}]).\label{18}
\eea
The above treatment is different from a master equation and solves the dynamics in a non-perturbative manner for the coupling. We derive the probability density of the position $x$ at time $t$
\bea
p(x,t)
&=&\Tr \bl[\ket{x}\bra{x}e^{-i\hat{H}_{\rm{p}}t}\h{\rho}^{I}e^{i\hat{H}_{\rm{p}}t} \br] \nnm \\
&=&A^{2}\int dx'dx'' \bl( \psi_{+}(x') + \psi_{-}(x') \br)\bl( \psi^{*}_{+}(x'') + \psi^{*}_{-}(x'') \br) \nnm \\
&\times& G(x-x',t)G^{*}(x-x'',t)  \exp \Bl[-W(x'-x'',T) \Br]  \nnm \\
&=:& \sum_{\al,\beta=\pm} p_{\al \beta}(x,t),
\eea
where $p_{\al \beta}(x,t)$ is defined as 
\bea
p_{\al \beta}(x,t)
&:=& A^{2}\int dx'dx''  \psi_{\al}(x')\psi^{*}_{\beta}(x'') G(x-x',t)G^{*}(x-x'',t) \exp \Bl[-W(x'-x'',T) \Br] \nnm \\
&=&\fr{A^{2}m_{\rm{p}}}{2\pi^{3/2}t}\int dx'dx''  \exp \Bl[-\fr{(x'-\al L)^{2}}{2}-i\fr{(x-x')^{2}}{2t/m_{\rm{p}}}\Br] \nnm \\
&\times& \exp \Bl[-\fr{(x''-\beta L)^{2}}{2}+i\fr{(x-x'')^{2}}{2t/m_{\rm{p}}}\Br]\exp \Bl[-W(x'-x'',T) \Br] \label{22}.
\eea
When $1/m$ (the ratio of the Compton length to the size of the slit) is sufficiently large for fixed $gT$, $mT$ and $2mL$, we can approximate the above equation (\ref{22}) as
\beq
p_{\al \beta}(x,t)
\approx \fr{A^{2}}{\sq{\pi(1+(t/m_{\rm{p}})^{2})}}\exp \Bl[-W(|\al-\beta|L,T)\Br] \exp \Bl[-\fr{(x-\al L)^{2}}{2(1+it/m_{\rm{p}})}-\fr{(x-\beta L)^{2}}{2(1-it/m_{\rm{p}})}\Br],
\label{23}
\eeq
for the large time $t/m_{\rm{p}} \gg 1$. The more detail of the derivation of the  equation (\ref{23}) is written in App.D. The probability density $p(x,t)$ is given as
\beq
p(x,t)
\approx A^{2} \Bl(|\psi_{+}(x,t)|^{2}+|\psi_{-}(x,t)|^{2}+e^{-W(2L,T)}[\psi_{+}^{*}(x,t)\psi_{-}(x,t)+\psi_{+}(x,t)\psi_{-}^{*}(x,t)]\Br).\label{24}
\eeq
For the above approximation, we use the conditions $T/m_{\rm{p}} \ll 1, 1/m \gg 1$~and $2L \gg 1$ to get the probability distribution (\ref{24}). These conditions describe that the particle is sufficiently localized than the fluctuation of the scalar field during the time evolution (the WKB approximation).
    
FIG. \ref{fig4} gives $p/A^{2}$ as a function of $x$ for $L=10, m=0.05,T=20,g=0.15$ and $t/m_{\rm{p}}=0,20 $.  We observe that the contrast of the interference fringes is reduced in the right panel of FIG.\ref{fig4}. In the next section, we introduce the visibility and FI matrix. Then a relation between them is derived.
\bfg[htbp]
    \incg[clip,height=4.5cm,width=14.0cm]{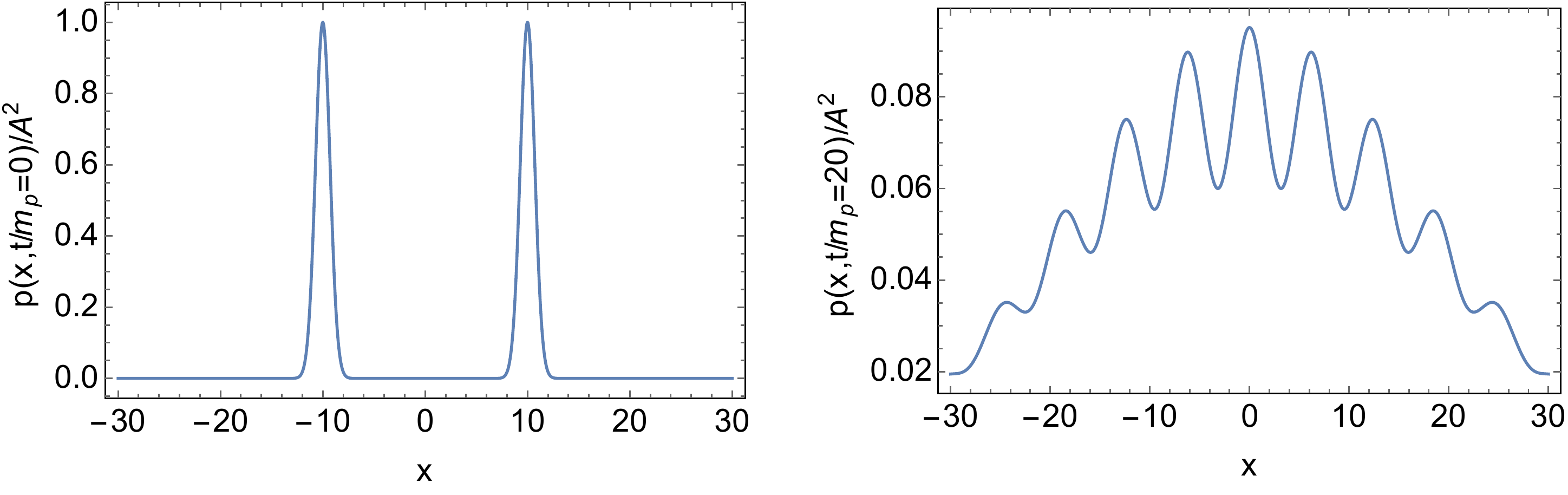}
    \caption{The behavior of $p/A^{2}$ as a function of $x$ for  $ L=10, m=0.05,T=20,g=0.15$ and $t/m_{p}=0,20 $. In the right panel the quantum coherence of the particle is reduced by the massive scalar field.}
    \label{fig4}
\efg

\section{The relation between the visibility and the Fisher information matrix}
Let us introduce the interferometric visibility to quantify the degree of the interference for the toy model of the double-slit experiment. It is given by 
\beq
{\mcl{V}}(t)=\fr{p(0,t)-p(x_{*},t)}{p(0,t)+p(x_{*},t)},
\eeq
where $x_{*}=\pi(1+(t/m_{\rm{p}})^{2})/2L(t/m_{\rm{p}})$ corresponds to the first minimum for $x \geq0$ of the oscillation term of the probability density. For $1 \ll 2L \ll t/m_{\rm{p}}$ and $t/m_{\rm{p}} \rightarrow \infty$, we derive 
\beq
{\mcl{V}}(t) \rightarrow \fr{(1+e^{-W})-e^{-\fr{\pi^{2}}{4L^{2}}}(1-e^{-W})}{(1+e^{-W})+e^{-\fr{\pi^{2}}{4L^{2}}}(1-e^{-W})}\approx e^{-W}. \label{26}
\eeq
The condition $2L\ll t/m_{\rm{p}}$ corresponds that the screen is distant from the double slit, and then the individual particles form the fringe pattern. Note that the visibility $\mcl{V}$ is equivalent to the coefficient of the interference term of the probability density $p(x,t)$. 

The probability distribution of the particle (\ref{24}) depends on the mass $m$ of the scalar field and the coupling coefficient $g$. Hence we can estimate the parameters and evaluate the estimation efficiency from the probability density. As well known in estimation theory, FI matrix\cite{13,14} characterizes an estimation precision of unknown parameters. In the next paragraph, let us give a brief overview of estimation theory and FI matrix. 

In estimation theory\cite{13}, we treat a probability distribution depending on unknown parameters. The probability distribution is denoted as a conditional probability $p(x|\tht)$, where $x$ is a measurement result, $\tht=(\tht_{1},\cdots,\tht_{i},\cdots,\tht_{n})^{T}$ and $\tht_{i}$ is an unknown parameter. In quantum mechanics, $p(x|\tht)$ is given by $\Tr [\Pi_{x}\rho_{\tht}]$, where $\rho_{\tht}$ is a density matrix and $\Pi_{x}$ is the elements of a positive operator-valued measure (POVM). To guess the true value of each parameter, we define a function $\tht_{i}(x)$ of the measurement result $x$, which is called an estimator of the parameter $\tht_{i}$. The estimator is unbiased if and only if the expectation value of the  estimator is equal to the true value of the parameter: 
\beq
\bkt{\tht_{i}}=\tht_{i},~\bkt{\tht_{i}}=\int dx~ \tht_{i}(x)p(x|\tht).
\eeq
To quantify an error of the estimation, we use a covariance matrix given as 
\beq
{C}_{ij}(\tht)=\int dx~ (\tht_{i}(x)-\bkt{\tht_{i}})(\tht_{j}(x)-\bkt{\tht_{j}})p(x|\tht). \label{29}
\eeq
If the eigenvalues of the covariance matrix $C(\tht)$ are small, then we estimate the parameters well. The covariance matrix depends on the choice of estimators. For any unbiased estimators, however, FI matrix defined as 
\beq
{\mcl{F}}_{ij}=\int dx~p(x|\tht)\Bl(\pa_{\tht_{i}}\ln p(x|\tht)\Br)\Bl(\pa_{\tht_{j}}\ln p(x|\tht)\Br) \label{30}
\eeq
gives a simple lower bound of the covariance matrix, which is known as the Cramer-Rao inequality\cite{14}:   
\beq
{C}(\tht) \geq  \mcl{F}^{-1} \label{31}.
\eeq
According to the Cramer-Rao inequality, if the eigenvalues of FI matrix are large then we can estimate the unknown parameters well. Hence FI matrix describes the estimation precision of the unbiased estimators. For the estimation problem of one parameter $\tht$, the Cramer-Rao inequality is given as 
\beq
{V}(\tht) \geq  \fr{1}{F(\tht)},\label{32}
\eeq
where the variance ${V}(\tht)$ and $F(\tht)$ are given by
\bea
{V}(\tht)&=&\int dx~ (\tht(x)-\bkt{\tht})(\tht(x)-\bkt{\tht})p(x|\tht), \nnm \\
F(\tht)&=&\int dx~p(x|\tht)\Bl(\pa_{\tht}\ln p(x|\tht)\Br)\Bl(\pa_{\tht}\ln p(x|\tht)\Br),\label{33}
\eea
where $F(\tht)$ is called FI of the parameter $\tht$. These definitions of the variance $V(\tht)$ and the Fisher information $F(\tht)$ correspond to each definition (\ref{29}) and (\ref{30}) for $n=1$. The Cramer-Rao inequality for one parameter leads to a classical signal-to-noise bound
\beq
R(\tht) \leq  \tht^{2}F(\tht), \label{34}
\eeq
where $R(\tht):=\tht^{2} /{V}(\tht) $ called the signal-to-noise ratio. In Sec. V we investigate the behavior of the classical and quantum signal-to-noise bound of the field mass or coupling. 

In the remaining of this section we calculate FI matrix  of the probability density (\ref{24}) and derive a relation to the interference visibility (\ref{26}) obtained from the distant screen $2L \ll t/m_{\rm{p}}$. Applying the above definition of FI matrix to our model, we can calculate FI matrix of the parameters $m$ and $g$ from the probability density (\ref{24}) as
\bea
\mcl{F}_{ij}(t) &=& \int dx~p(x,t)\pa_{\tht_{i}}\ln p(x,t) \pa_{\tht_{j}}\ln p(x,t) \nnm \\
&=&\fr{A^{2}\sq{1+(t/m_{\rm{p}})^{2}}}{\sq{\pi} Lt/m_{\rm{p}}}\exp \Bl[-\fr{L^{2}}{1+(t/m_{\rm{p}})^{2}}-2W \Br]\Bl(\pa_{\tht_{i}}W\pa_{\tht_{j}}W\Br) \nnm \\
&\times&\int du\fr{\cos^{2}u}{\cosh \bl(\fr{u}{t/m_{\rm{p}}}\br)+e^{-W}\cos u}\exp \Br[-\fr{1+(t/m_{\rm{p}})^{2}}{4L^{2}(t/m_{\rm{p}})^{2}}u^{2}\Br] \label{35}
\eea
where $\tht_{1}=m$ and $\tht_{2}=g$. In the second line of (\ref{35}), we have substituted $p(x,t)$ into the definition of FI matrix, and the integral variable $x$ has been changed as $u=2Lx(t/m_{\rm{p}})/(1+(t/m_{\rm{p}})^{2})$. In the limit $t/m_{\rm{p}} \rightarrow \infty$, we get the integral as follows:
\bea
\mcl{F}_{ij}(t)&\rightarrow& \fr{A^{2}}{\sq{\pi}L}e^{-2W}\Bl(\pa_{\tht_{i}}W\pa_{\tht_{j}}W\Br)\int du\fr{\cos^{2}u}{1+e^{-W}\cos u} \exp \Br[-\fr{u^{2}}{4L^{2}}\Br]\nnm \\
&=&\fr{A^{2}}{\sq{\pi}L}\Bl(\pa_{\tht_{i}}W\pa_{\tht_{j}}W\Br)\Bl(\int du\fr{\exp \bl[-\fr{u^{2}}{4L^{2}}\br]}{1+e^{-W}\cos u} +2\sq{\pi}L(e^{-W-L^{2}}-1)\Br) . \label{36}
\eea
For $2L \gg 1$ we derive the following analytic form of FI matrix by the function $W$ or the visibility $\mcl{V}$ :
\bea
\mcl{F}_{ij}(t)
&\approx& \Bl(\fr{1}{\sq{1-e^{-2W}}}-1 \Br)\Bl(\pa_{\tht_{i}}W\pa_{\tht_{j}}W\Br) \nnm \\
&=& \Bl(\fr{1}{\sq{1-\mcl{V}^{2}}}-1 \Br)\Bl(\pa_{\tht_{i}}\log \mcl{V}\Br)\Bl(\pa_{\tht_{j}}\log \mcl{V}\Br). \label{37}
\eea
The calculation of the integral in (\ref{36}) is written in the App.E. The equation (\ref{37}) denote that FI matrix has no inverse. Due to the Cramer-Rao bound (\ref{31}), we cannot estimate the two parameters simultaneously with high precision. This is because there is the parameter dependence only in the contrast of the fringe pattern and the estimated parameters degenerate. In the next section, we show the explicit function of the visibility and FI of each parameter.


\section{The behavior of the visibility and the Fisher information}
We have derived the relation between the visibility $\mcl{V}$ and FI matrix $\mcl{F}_{ij}$ of the mass $m$ and the coupling $g$ in the previous section. The relation (\ref{37}) show that FI matrix given by the visibility have no inverse, that is we cannot estimate the two parameters simultaneously. Hence we consider the estimation precision of each parameter (the diagonal components of FI) in this section. 

To treat the visibility and FI analytically, we consider the case where the interaction time $T$ satisfies the conditions $mT \gg 1$ and $T \gg 2L$. This case corresponds to that the scalar field is exchanged many times in each path or between two paths (FIG.\ref{fig5-1}). 
\bfg[htbp]
 \bmn{0.5\hsize}
  \incg[clip,height=6.0cm,width=5.0cm]{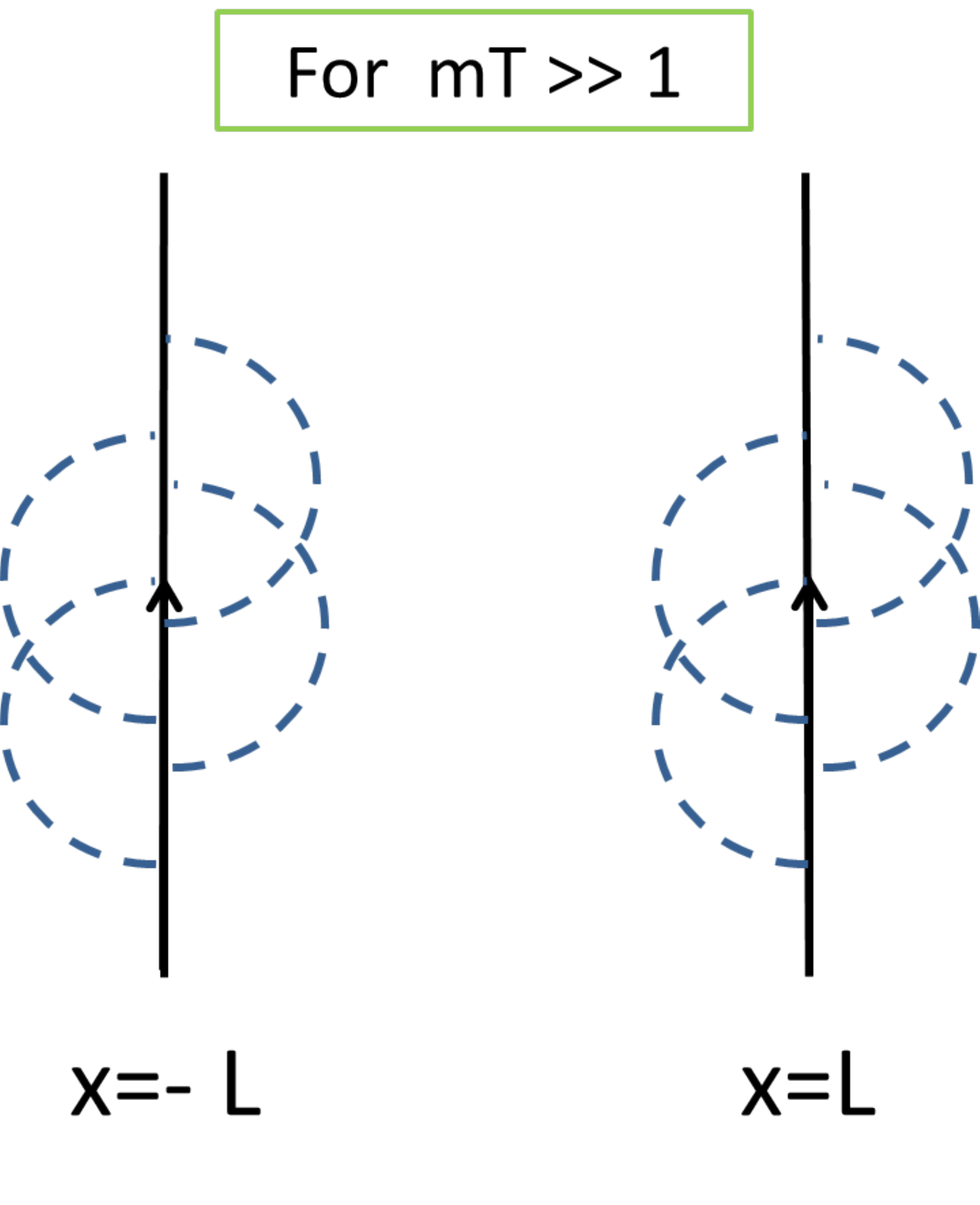}
 \emn
 \bmn{0.5\hsize}
  \incg[clip,height=6.0cm,width=5.0cm]{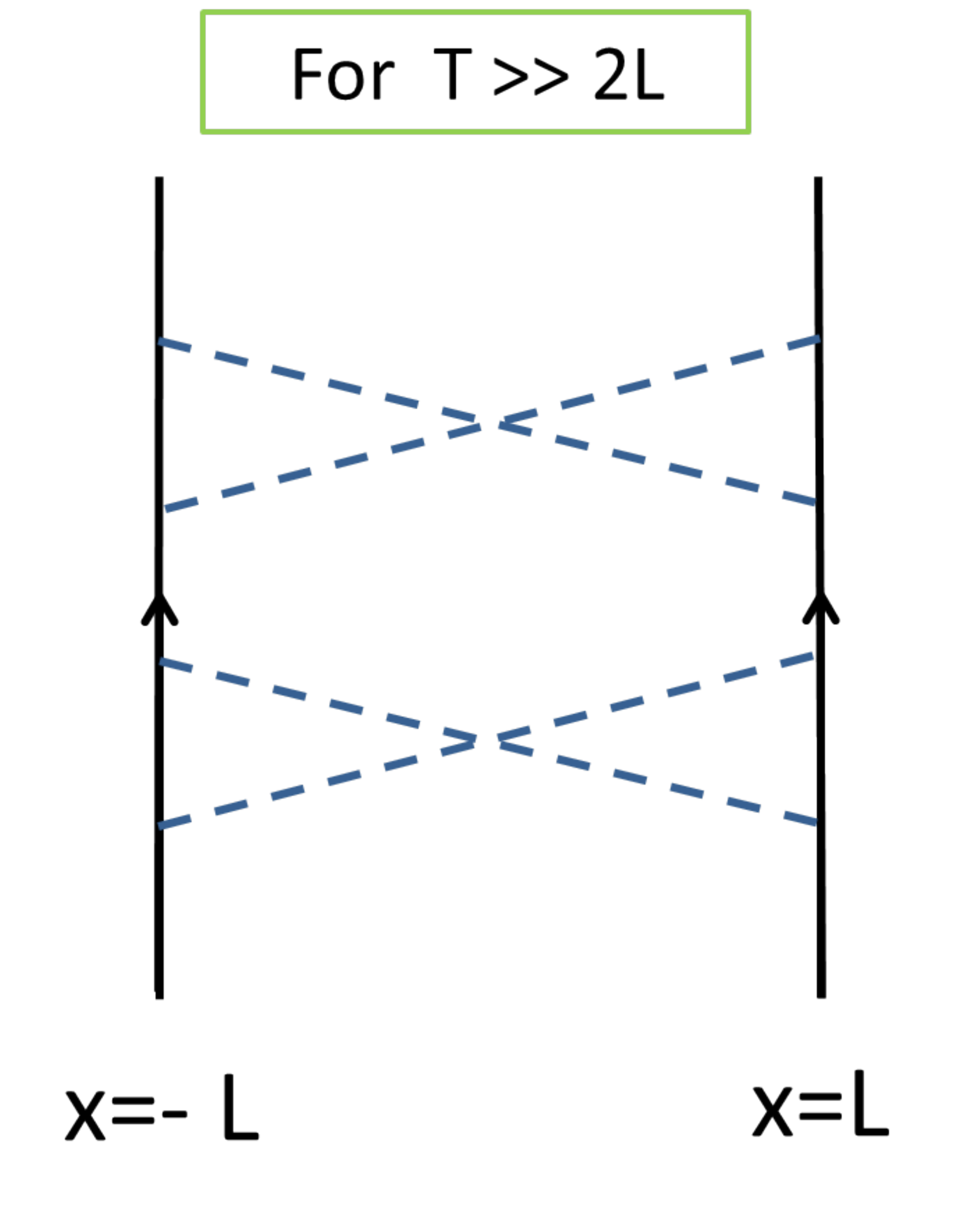}
 \emn
    \caption{The schematic diagrams of the contributions included in the function $W(2L,T)$ for $mT \gg 1$ and $T \gg 2L$. In this case, the scalar field is exchanged many times in each path or between two paths. }
    \label{fig5-1}
\efg
For $mT \gg 1$ and $T \gg 2L$ the function $W(2L,T)$ is calculated approximately by the Riemann-Lebesgue lemma:
\bea
W(2L,T)
&=&\fr{g^{2}}{\pi m^{2}}\int^{\infty}_{0} \fr{du}{(u^{2}+1)^{3/2}}(1-\cos[2mLu])(1-\cos[mT\sq{u^{2}+1}]) \nnm \\
&\approx& \fr{g^{2}}{\pi m^{2}}\int^{\infty}_{0} \fr{du}{(u^{2}+1)^{3/2}}(1-\cos[2mLu]) \nnm \\
&=&\fr{g^{2}}{\pi m^{2}}\Bl(1-2mL K_{1}(2mL)\Br),
\eea
where $K_{\nu}(z)$ is the modified Bessel function. The visibility $\mcl{V}$ is given as
\bea
\mcl{V}=\exp \Bl[-\fr{g^{2}}{\pi m^{2}} \Bl(1-2mL K_{1}(2mL) \Br) \Br].
\eea
The classical bound of the signal-to-noise ratio of each parameter is given as 
\bea
m^{2}F(m)&=&\Bl \{ \Bl( 1-e^{-\fr{2g^{2}}{\pi m^{2}} (1-2mL K_{1}(2mL) ) } \Br)^{-\frac{1}{2}}-1 \Br \} \Bl[\fr{2g^{2}}{\pi m^{2}} \bl(1-2(mL)^{2}K_{2}(2mL)\br)\Br]^{2} , \nnm \\ 
g^{2}F(g)&=&\Bl \{ \Bl( 1-e^{-\fr{2g^{2}}{\pi m^{2}} (1-2mL K_{1}(2mL) ) } \Br)^{-\frac{1}{2}}-1 \Br \}\Bl[\fr{2g^{2}}{\pi m^{2}} (1-2mL K_{1}(2mL))\Br]^{2}.
\eea
FIG. \ref{fig5} shows $\mcl{V}$, $m^{2}F(m)$ and $g^{2}F(g)$ as a function $2mL$ for fixed $g/m=1.5$. 
\bfg[htbp]
 \bmn{0.5\hsize}
  \incg[clip,height=4.0cm,width=6.5cm]{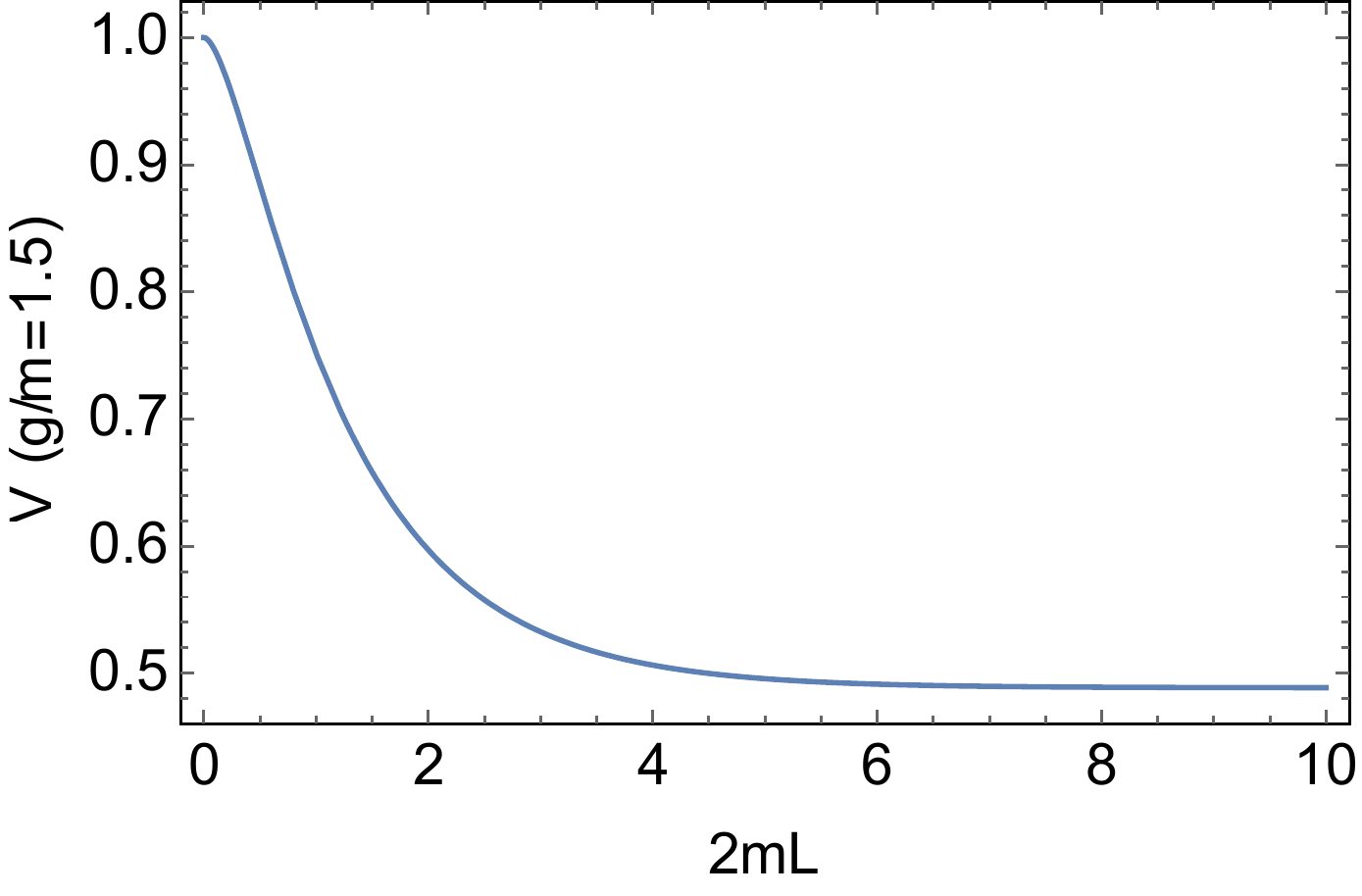}
 \emn
 \bmn{0.5\hsize}
  \incg[clip,height=4.0cm,width=8.0cm]{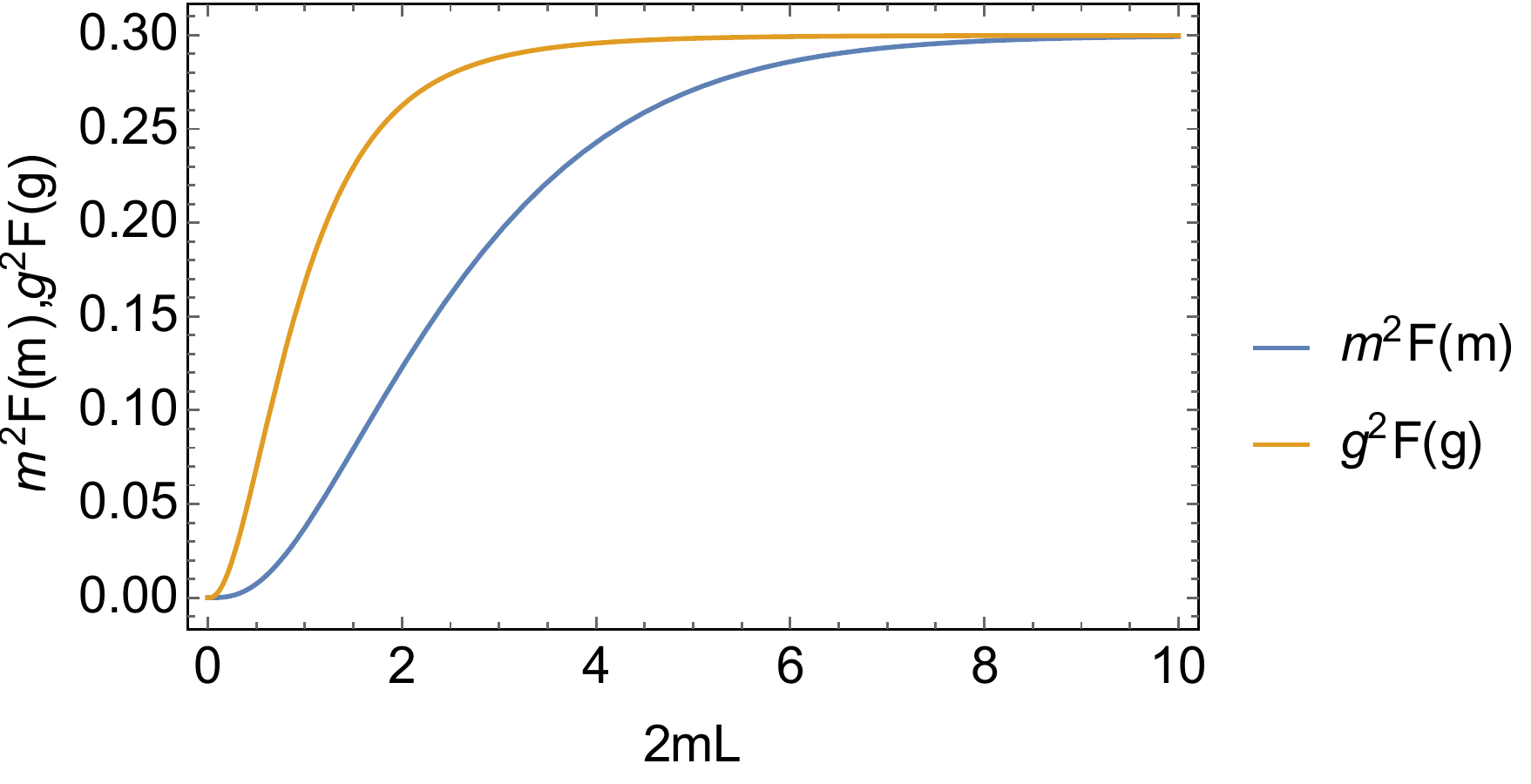}
 \emn
    \caption{The visibility and the classical bounds as a function of $2mL$ for $g/m=1.5$. }
    \label{fig5}
\efg
In FIG. \ref{fig5} the visibility decreases and the classical bounds (the Fisher informations) increase as $2mL$ becomes large. These behaviors are understood as follows: the vacuum of the scalar field is excited by the interaction with the particle which have passed through the slit. The excited state differs depending on which slit the particle goes through (in the first panel of FIG. \ref{fig6}). When $2mL \ll 1$ the excited state on each path degenerates (in the second panel of FIG. \ref{fig6}). On the other hand, when $2mL \gg1$, the excited state is distinguished well (in the third panel of FIG. \ref{fig6}). 
\bfg[htbp]
 \bmn{0.33\hsize}
  \incg[clip,height=4.0cm,width=5.4cm]{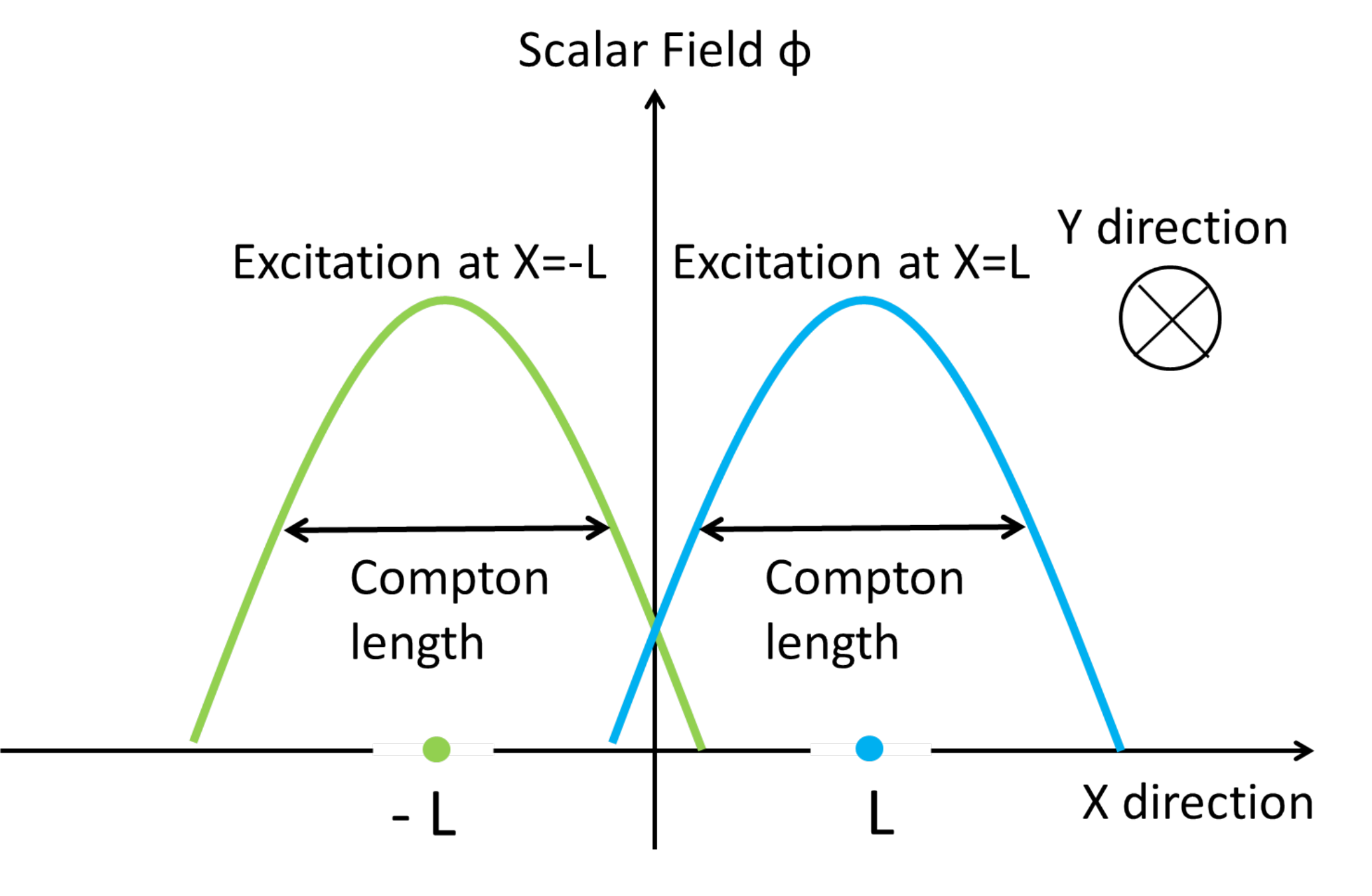}
 \emn
 \bmn{0.33\hsize}
  \incg[clip,height=4.0cm,width=5.4cm]{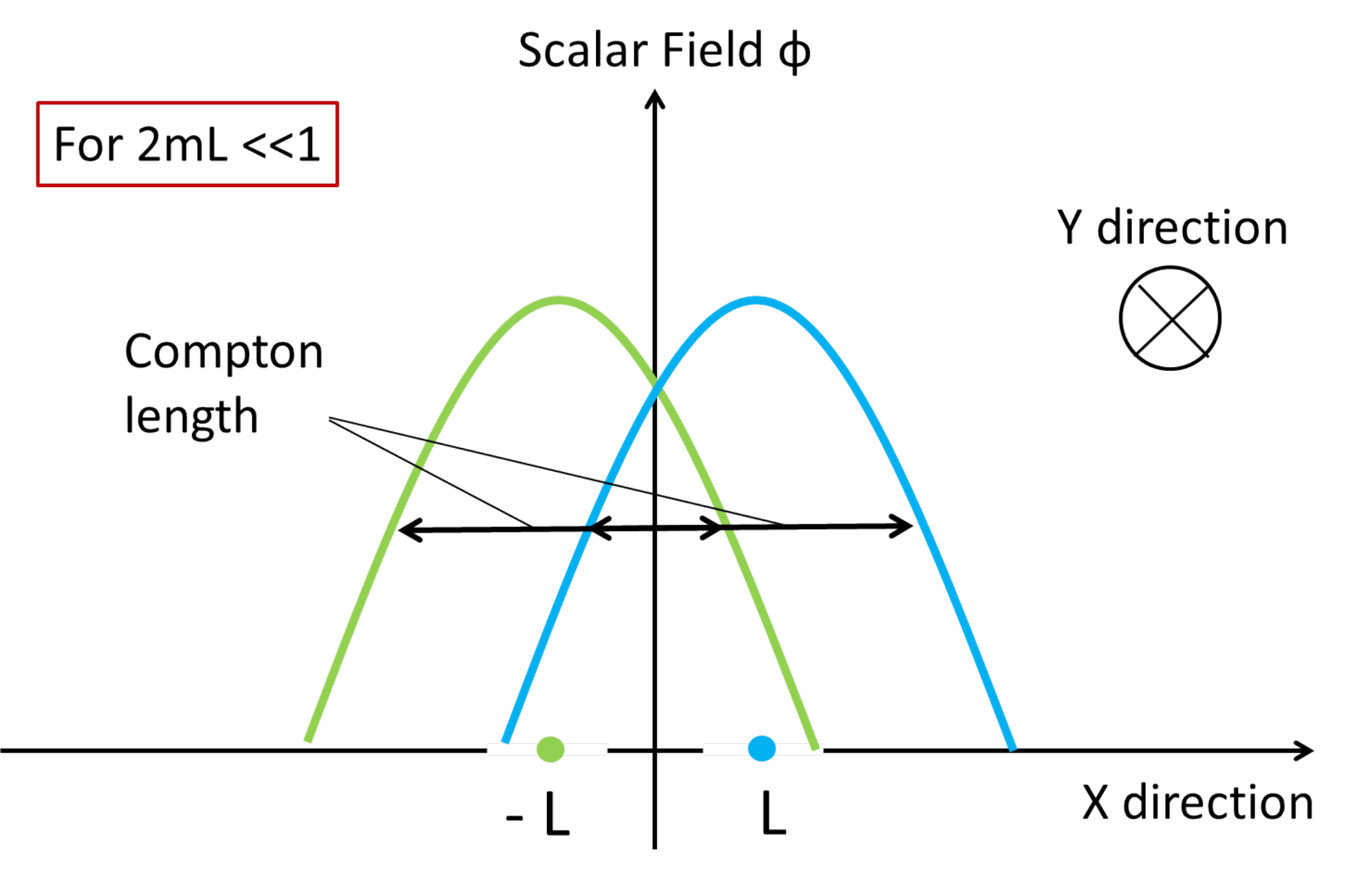}
 \emn
 \bmn{0.33\hsize}
  \incg[clip,height=4.0cm,width=5.4cm]{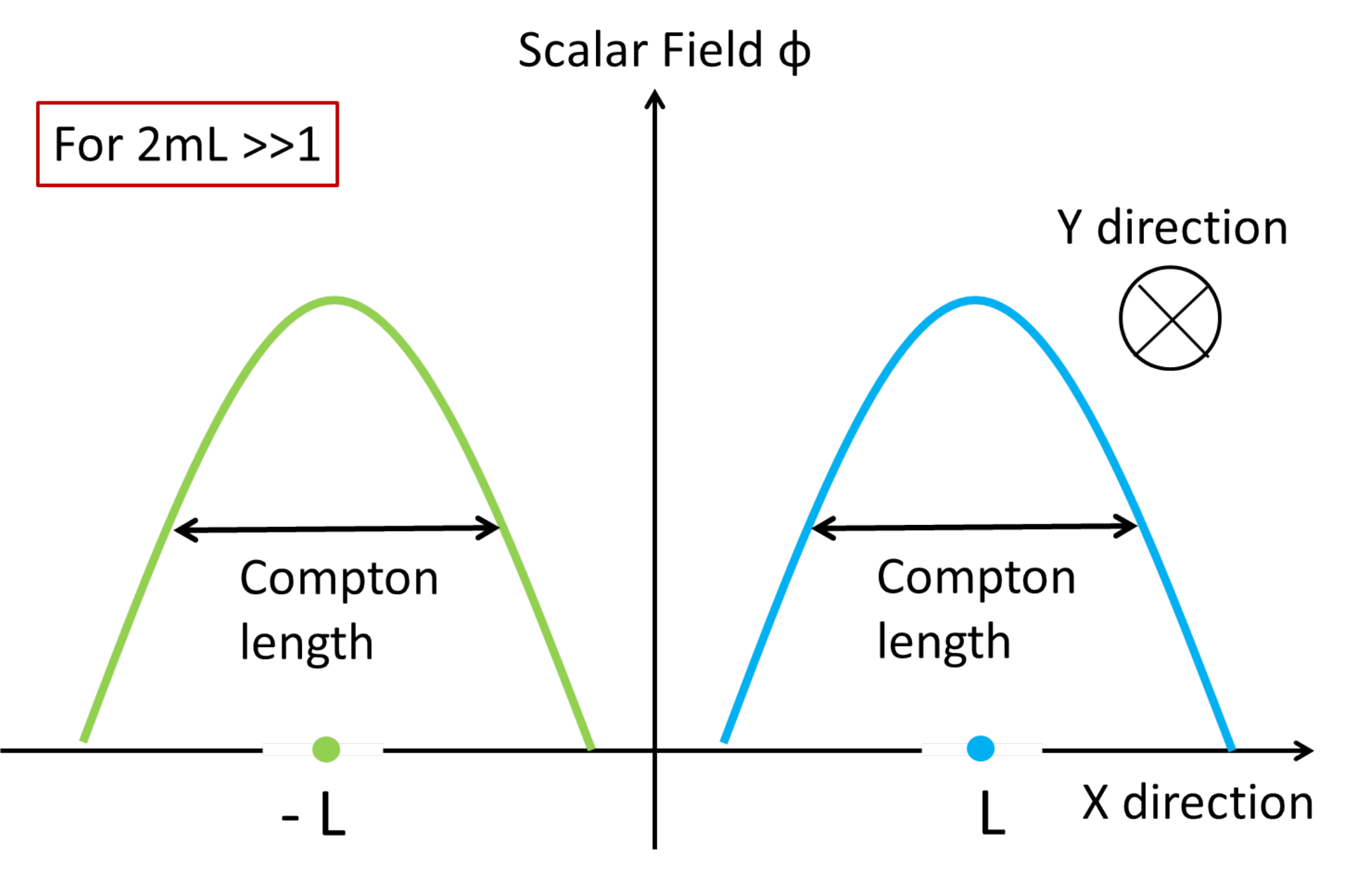}
 \emn
    \caption{The excited state of the scalar field differs depending on the path of the particle. The green and blue curve describe the schematic diagrams of the excitation at $x=-L$ and $x=L$. Each of the excited states has the width of the Compton wave length. In the second figure (for $2mL \ll1$) the scalar field cannot detect the path of the particle well. However, in the third figure (for $2mL \gg 1$) each of the states has no correlation. Hence the scalar field can distinguish which path the particle has passed through.}
    \label{fig6}
\efg
That is, the massive scalar field can be regarded as a detector which extracts the which-way information \cite{27,28}. Hence the massive scalar field detects the path of the particle and induces the quantum decoherence. According with the loss of the coherence by the massive scalar field, the particle carries the information of the field parameter. This leads to the gain of FI as seen in the right panel of FIG.\ref{fig5}. In the limit $2mL \rightarrow \infty$, $\mcl{V}$, $m^{2}F(m)$ and $g^{2}F(g)$ are given as    
\bea
\mcl{V}\rightarrow \exp \Bl[-\fr{g^{2}}{\pi m^{2}}\Br],~~~m^{2}F(m),g^{2}F(g)\rightarrow \Bl \{ \Bl(1-\exp \Bl[-\dip{\fr{2g^{2}}{\pi m^{2}}} \Br] \Br)^{-\frac{1}{2}}-1 \Br \} \Bl[\fr{2g^{2}}{\pi m^{2}} \Br]^{2} \label{42}.
\eea
Due to the fact that we assume the large time limit $t/m_{p} \rightarrow \infty$ (the screen is far from the double slit), the particle interferes certainly and the visibility remains in the limit $2mL\rightarrow \infty$. Similarly, the classical bounds $m^{2}F(m)$ and $g^{2}F(g)$ derived from the Cramer-Rao inequality (\ref{34}) do not vanish in the limit $2mL \rightarrow \infty$. The two functions approach the above same value. This is because the estimated parameters degenerate in the fringe pattern and the visibility depends only on the ratio $g/m$ for $mT \gg 2mL \gg1$. 

As mentioned above, we have derived that the estimation precision is improved by wave nature of the particle (the right panel of FIG. \ref{fig5}). In the recent works \cite{32,33},      a particle's mass estimation problems in a gravitational field background is discussed, and it was shown that the mass estimation precision gets better when the spread of the particle's wave packet is comparable with the curvature scale. The previous results are similar to our calculation and the points of difference are considered as follows: in our model, we apply the WKB approximation to solve the dynamics of a particle coupled to a massive scalar field. If we consider a single path of the particle (single wave packet)  then the wave function of the particle is unchanged and independent of the coupling. However, we have considered that the particle's wave function is the superposition of two wave packets (that is, a double path of the particle). If the interval between each peak of the two packets is greater than the Compton length of the scalar field (``the curvature scale" in our model), then the decoherence occurs. Hence we conclude that the parameter estimation in our model is based on not the spread of a wave function, but the reduction of wave interference effects. 

Finally, let us assume that the coupling constant $g$ is controllable and focus on the classical bound $m^{2}F(m)$ of the mass in the equation (\ref{42}). FIG.\ref{fig7} presents $m^{2}F(m)$ as a function $g/m$ for fixed $m$. 
\bfg[htbp]
  \incg[clip,height=5.0cm,width=8.0cm]{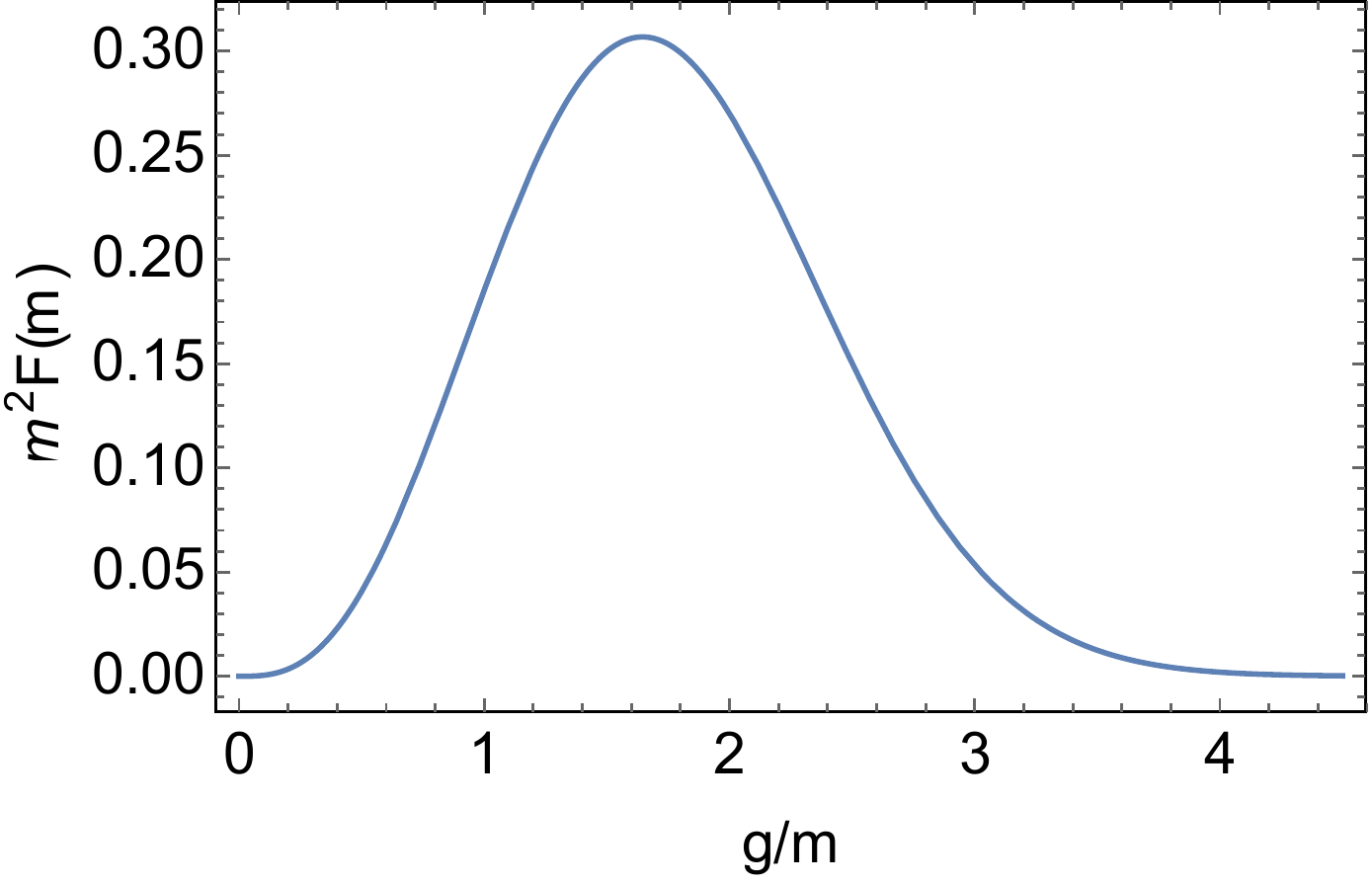}
    \caption{The classical bound $m^{2}F(m)$ of the field mass as a function $g/m$ for $2mL \rightarrow \infty$ and fixed $m$. There is a peak at $g/m \simeq O(1)$.}
    \label{fig7}
\efg
The behavior of the classical bound $m^{2}F(m)$ is explained as follows: when the coupling $g$ is small, it is difficult to induce the quantum decoherence and estimate the mass parameter. When the coupling $g$ is large,the superposition of the particle which passes through the double slit disappears. However, the distribution on the screen has no information about the massive scalar field because the particle just goes through the each slit with a certain classical probability in this case. Thus, we conclude that FI given by the quantum decoherence in the double-slit interferometer characterizes the intermediate behavior of wave propagation and particle motion. 
\section{Conclusion}
We have investigated the estimation problem of the parameters of a massive scalar field by the quantum decoherence observed in the double-slit interferometer. In the previous works \cite{14,15,18} of quantum probe, the dynamics of the probe was treated by the quantum master equation. However, we have solved the time evolution of the particle interacting with the scalar field in the non-perturbative treatment (the WKB approximation). This formalism provides a method to analyze the estimation problem for the case of large coupling. To quantify the degree of quantum interference effect and the estimation precision of the mass and coupling, we have computed the interference visibility and FI matrix of the parameters. The visibility describes the coefficient of the interference term and FI matrix gives a lower bound of the covariance matrix of estimated parameters in the sense of the Cramer-Rao inequality. 

In the large time limit (the position of the screen is sufficiently far) and for the large distance between the slits, we have derived the explicit relation between the visibility and FI matrix of the mass and the coupling. The simple relation gives two main properties: FI matrix is calculated from the visibility given by only the difference of the probability density at the origin ($x=0$) and the point of the first minimum ($x=x_{*}$) on the screen. FI matrix has no inverse because the estimated parameters, mass and coupling, degenerate in the fringe pattern. The first property gives a simple method to evaluate the estimation error in the double-slit experiment, and the second one tells us that the quantum decoherence is not useful to estimate multiple parameters. Here, we comment on the comparison between our calculation and the previous works. In \cite{29,30,31}, the parameter estimation using the phase shift in a matter-wave interferometer was discussed and FI for a position measurement was calculated. On the other hand, above FI matrix given by the equation (\ref{37}) is obtained from not the phase shift but the contrast of the interference fringes. Hence the simple relation derived in this paper is novel. 

We have focused on the behavior of the visibility and FI of each parameter. When the distance between the two slits is much larger than the Compton length of the field, the visibility becomes small. According with that, the classical bound given by FI has a large value. This is because the which-way information extracted by the scalar field increases when the distance between the two slit is much greater than the Compton wave length. 

We have presented FI of the field mass as a function of the coupling in Sec.V. When the coupling parameter is small, it is difficult for the probability distribution of the particle to depend on the field mass. Then the interference pattern does not vanish and FI of the mass parameter is small. For large coupling, the probability density of the particle has no interference term in our approximation. The particle follows a classical  distribution which is independent of the field mass. Hence we also have observed that FI becomes small. Thus the peak position of FI of the mass denotes an intermediate situation between wave and particle nature. This property implies that it is possible to quantify the wave-particle duality by evaluating FI obtained from the quantum decoherence. The relation between quantum-to-classical transition and FI given by quantum decoherence may be revealed in the future study.
\section*{Acknoledgement}
We thank Chul-Moon Yoo for useful discussion.
\appendix
{\flushleft{\section{A non-relativistic particle motion : reduction to a one-dimensional quantum system}}}
We consider the free Hamiltonian of a non-relativistic particle 
\beq
\h{H}=\fr{1}{2m_{\rm{p}}} \Bl(\h{p}^{2}_{x}+\h{p}^{2}_{y}+\h{p}^{2}_{z}\Br), 
\eeq
where $\h{p}_{x},\h{p}_{y}$ and $\h{p}_{z}$ are canonical momenta conjugate to positions of the particle $x,y$ and $z$, respectively. $m_{\rm{p}}$ is the mass of the particle. The canonical variables obey the canonical commutation relations 
\beq
\bl[x,p_{x}\br]=\bl[y,p_{y}\br]=\bl[z,p_{z}\br]=i,~others=0. 
\eeq
In a double-slit experiment, a detection screen catches the particles which have passed through the double slit. We assume that the $x,y$ and $z$ axes are as shown in the FIG.\ref{fig1} and the position of the screen is $y=y_{\rm{sc}}$. A histogram of the particle on the screen at an observation time $t$ is given by
\beq
p_{\rm{sc}}(x,t)=\dip{\fr{\dip{\int} dz~p(x,y=y_{\rm{sc}},z,t)}{\dip{\int} dxdz~p(x,y=y_{\rm{sc}},z,t)}}, 
\eeq
where the probability density $p(x,y,z,t)$ is the square of the absolute value of the wave function $\psi(x,y,z,t)$. The definition of the wave function is   
\beq
\psi(x,y,z,t):=\bra{x,y,z}e^{-i\h{H}_{p}t} \ket{\psi}, \label{A4}
\eeq
where $\ket{x,y,z}$ is the position eigenstates and $\ket{\psi}$ is an initial state. In particular, if the initial wave function $\psi(x,y,z)=\bkt{x,y,z|\psi}$ is a separable function, that is, 
\beq
\psi(x,y,z)=\psi_{1}(x)\psi_{2}(y)\psi_{3}(z), 
\eeq
then we find that $\psi(x,y,z,t)$ has a separable form as 
\beq
\psi(x,y,z,t)=\psi_{1}(x,t)\psi_{2}(y,t)\psi_{3}(z,t) 
\eeq
from the equation (\ref{A4}). Using the separable form of the wave function $\psi(x,y,z,t)$, we can get the simple formula of $p_{\rm{sc}}(x,t)$ as follows: 
\beq
p_{\rm{sc}}(x,t)=\dip{\fr{\dip{\int} dz~p(x,y=y_{\rm{sc}},z,t)}{\dip{\int} dxdz~p(x,y=y_{\rm{sc}},z,t)}} =\dip{\fr{\dip{\int} dz~p_{1}(x,t)p_{2}(y_{\rm{sc}},t)p_{3}(z,t)}{\dip{\int} dxdz~p_{1}(x,t)p_{2}(y_{\rm{sc}},t)p_{3}(z,t)}}=\dip{\fr{p_{1}(x,t)}{\dip{\int} dx~p_{1}(x,t)}}.  
\eeq
Note that $\psi_{1}(x,t)$ obeys the one-dimensional Schr\"{o}dinger equation. Hence, the probability is conserved and we can obtain the more simple form of $p_{\rm{sc}}$ as
\beq
p_{\rm{sc}}(x,t)=\fr{1}{N}p_{1}(x,t), 
\eeq
where $N$ is a normalization factor. Therefore we can describe the double-slit experiment as the one-dimensional system under the above initial condition.
{\flushleft{\section{Approximation in the equation (\ref{13})}}}
To show the explicit calculation in the equation (\ref{13}), we first consider the initial state 
\beq
\ket{\Psi_{\rm{in}}}=\ket{\psi}\ket{0_{\phi}}, \psi(x)=\fr{1}{\pi^{1/4}}\exp \Bl[-\fr{x^{2}}{2} \Br], 
\eeq
where $\psi(x)=\bkt{x|\psi}$. The time evolution is given by
\bea
\ket{\Psi_{\rm{in}}}=\ket{\psi}\ket{0_{\phi}}
&\longrightarrow&{\rm{T_{p} \exp}}\bl[-i\int^{T}_{0} dt H_{int}(t) \br]\ket{\psi}\ket{0_{\phi}} \nnm \\
&=&{\rm{T_{p} \exp}}\bl[-ig\int^{T}_{0} dt\int dx \dlt(x-\h{x}(t)) \h{\phi}(x,t) \br]\ket{\psi}\ket{0_{\phi}} \nnm \\
&=&\sum_{n=0}^{\infty}\fr{(-ig)^{n}}{n!}\int dx_{0}\Bl[\prod_{i=1}^{n}\int^{T}_{0} dt_{i}\int dx_{i}\Br] \int dx_{n+1} \ket{x_{0}}G(x_{0}-x_{1},-t_{1}) \nnm \\
&\times& \Bl[\prod_{i=1}^{n-1}G(x_{i}-x_{i+1},t_{i}-t_{i+1})\Br]G(x_{n}-x_{n+1},t_{n}) \psi(x_{n+1}) \nnm \\
&\times&{\rm{T_{p}}}\Bl(\h{\phi}(x_{1},t_{1})\cdots\h{\phi}(x_{n},t_{n})\Br)\ket{0_{\phi}}, 
\eea 
where $\ket{x}$ is the eigenstate of $\h{x}$. The function $G$ is derived as
\beq
G(x-x',t-t')=\sq{\fr{m_{\rm{p}}}{2\pi i (t-t')}}\exp \Bl[-\fr{m_{\rm{p}}(x-x')^{2}}{2i(t-t')} \Br]
\eeq
We define the variable as $\tau=t/T$. The above equation is rewritten as 
\bea
{\rm{T_{p} \exp}}\bl[-i\int^{T}_{0} dt H_{int}(t) \br]\ket{\psi}\ket{0_{\phi}} 
&=&\sum_{n=0}^{\infty}\fr{(-igT)^{n}}{n!}\int dx_{0}\Bl[\prod_{i=1}^{n}\int^{1}_{0} d\tau_{i}\int dx_{i}\Br] \int dx_{n+1} \ket{x_{0}} \nnm \\ 
&\times&\ti{G}(x_{0}-x_{1},-\tau_{1})\Bl[\prod_{i=1}^{n-1}\ti{G}(x_{i}-x_{i+1},\tau_{i}-\tau_{i+1})\Br]\ti{G}(x_{n}-x_{n+1},\tau_{n}) \nnm \\
&\times& \psi(x_{n+1}) {\rm{T_{p}}}\Bl(\h{\ti{\phi}}(x_{1},\tau_{1})\cdots\h{\ti{\phi}}(x_{n},\tau_{n})\Br)\ket{0_{\phi}}, 
\eea
where $\h{\ti{\phi}}(x,\tau)=\h{\phi}(x,t)$. The function $\ti{G}$ is  
\beq
\ti{G}(x-x',\tau-\tau')=\sq{\fr{1}{2\pi i (\tau-\tau')\eps^{2}}}\exp \Bl[-\fr{(x-x')^{2}}{2i\eps^{2}(\tau-\tau')} \Br],~\eps^{2}=\fr{T}{m_{\rm{p}}}. 
\eeq
If the initial wave packet is almost unchanged during the interaction time , that is, $\eps^{2}$ is sufficiently smaller than $1$, then $\ti{G}(x,\tau)$ is approximated by the delta function $\dlt(x)$. In this approximation, we get the following formula of the time evolution: 
\bea
{\rm{T_{p} \exp}}\bl[-i\int^{T}_{0} dt H_{int}(t) \br]\ket{\psi}\ket{0_{\phi}}
&\approx&\sum_{n=0}^{\infty}\fr{(-igT)^{n}}{n!}\int dx_{0}\Bl[\prod_{i=1}^{n}\int^{1}_{0} d\tau_{i}\int dx_{i}\Br] \int dx_{n+1} \ket{x_{0}} \nnm \\
&\times&\dlt(x_{0}-x_{1}) \Bl[\prod_{i=1}^{n-1}\dlt(x_{i}-x_{i+1})\Br]\dlt(x_{n}-x_{n+1}) \nnm \\
&\times&\psi(x_{n+1}) {\rm{T_{p}}}\Bl(\h{\ti{\phi}}(x_{1},\tau_{1})\cdots\h{\ti{\phi}}(x_{n},\tau_{n})\Br)\ket{0_{\phi}} \nnm\\
&=&\sum_{n=0}^{\infty}\fr{(-igT)^{n}}{n!}\int dx_{0}\Bl[\prod_{i=1}^{n}\int^{1}_{0} d\tau_{i}\Br] \psi(x_{0}) \ket{x_{0}} \nnm \\
&\times&{\rm{T_{p}}}\Bl(\h{\ti{\phi}}(x_{0},\tau_{1})\cdots\h{\ti{\phi}}(x_{0},\tau_{n})\Br)\ket{0_{\phi}} \nnm \\
&=&\int dx \psi(x) \ket{x} {\rm{T_{p}}} \exp \bl[-igT \int^{1}_{0} d\tau\h{\ti{\phi}}(x,\tau) \br] \ket{0_{\phi}}. \nnm \\
&=&\int dx \psi(x) \ket{x} {\rm{T_{p}}} \exp \bl[-ig \int^{T}_{0} dt\h{\phi}(x,t) \br] \ket{0_{\phi}}. 
\eea
The equation (\ref{13}) is obtained by applying the above approximation to the each Gaussian function.
{\flushleft{\section{Derivation of the equation (\ref{18})}}}
The massive scalar field $\h{\phi}(x,t)$ is given by 
\bea
\h{\phi}(x,t)=\int^{\infty}_{-\infty} \fr{dk}{\sq{4\pi \omg_{k}}} \Bl(\hat{a}_{k}e^{-i\omg_{k}t+ikx}+\hat{a}_{k}^{\dg}e^{i\omg_{k}t-ikx}\Br),
\eea
where $\omg_{k}=\sq{k^{2}+m^{2}}$. $\hat{a}_{k}$ is an annihilation operator, which satisfies the following commutation relations:
\bea
[\hat{a}_{k},~\hat{a}^{\dg}_{k'}]=\dlt(k-k'),~[\hat{a}_{k},~\hat{a}_{k'}]=[\hat{a}^{\dg}_{k},~\hat{a}^{\dg}_{k'}]=0.
\eea
The vacuum state $\ket{0_{\phi}}$ of the scalar field is defined as $\hat{a}_{k} \ket{0_{\phi}}=0$. The two point function $\bra{0_{\phi}}\hat{\phi}(x,t)\hat{\phi}(x',t')\ket{0_{\phi}}$ is calculated as
\bea
\bra{0_{\phi}}\h{\phi}(x,t)\h{\phi}(x',t')\ket{0_{\phi}}
&=&\int^{\infty}_{-\infty} \fr{dk}{4\pi \omg_{k}}e^{-i\omg_{k}(t-t')+ik(x-x')} \nnm \\
&=&\int^{\infty}_{-\infty} \fr{du}{4\pi \sq{u^{2}+1}}e^{-im(t-t')\sq{u^{2}+1}+im(x-x')u}, \label{C3}
\eea
where the integral variable $u=k/m$. We derive the explicit formula of the function $W$ as
\bea
W&=&\fr{g^{2}}{2}\int^{t}_{0} dt \int^{t}_{0} dt' \bra{0_{\phi}}\Bl({\rm{T_{p}}}(\h{\phi}(x,t)\h{\phi}(x,t'))^{\dg}+{\rm{T_{p}}}(\h{\phi}(x',t)\h{\phi}(x',t')-2\h{\phi}(x,t)\h{\phi}(x',t') \Br)\ket{0_{\phi}} \nnm \\
&=&\fr{g^{2}}{2}\int^{t}_{0} dt \int^{t}_{0} dt' \Bl(\tht(t-t')\bra{0_{\phi}}(\h{\phi}(x,t')\h{\phi}(x,t)+\h{\phi}(x',t)\h{\phi}(x',t'))\ket{0_{\phi}} \nnm \\
&+&\tht(t'-t)\bra{0_{\phi}}(\h{\phi}(x,t)\h{\phi}(x,t')+\h{\phi}(x',t')\h{\phi}(x',t))\ket{0_{\phi}} -2\bra{0_{\phi}}\h{\phi}(x,t)\h{\phi}(x',t')\ket{0_{\phi}} \Br) \nnm \\
&=&\fr{g^{2}}{\pi m^{2}}\int^{\infty}_{0} \fr{du}{(u^{2}+1)^{3/2}}(1-\cos[m(x-x')u])(1-\cos[mt\sq{u^{2}+1}]),
\eea
where we used the equation (\ref{C3}) for the last line of the above calculation.
{\flushleft{\section{Approximation of the integral (\ref{23})}}}
Let us explain the approximation of the function $p_{\al \beta}(\al, \beta=\pm)$ and the derivation of the condition of $W$.  $p_{\al \beta}(x,t)$ is  
\bea
p_{\al \beta}(x,t)
&=&\fr{A^{2}}{2\pi^{3/2}t/m_{\rm{p}}}\int dx'dx''  \exp \Bl[-\fr{(x'-\al L)^{2}}{2}-i\fr{(x-x')^{2}}{2t/m_{\rm{p}}}\Br] \nnm \\
&\times&\exp \Bl[-\fr{(x''-\beta L)^{2}}{2}+i\fr{(x-x'')^{2}}{2t/m_{\rm{p}}}\Br] \exp \bl[-W(x'-x'',T) \br] \nnm \\
&=&\fr{A^{2}}{2\pi^{3/2}t/m_{\rm{p}}}\int dx'dx''  \exp \Bl[-\fr{x'^{2}}{2}-i\fr{(x-\al L-x')^{2}}{2t/m_{\rm{p}}}\Br] \nnm \\
&\times& \exp \Bl[-\fr{x''^{2}}{2}+i\fr{(x-\beta L-x'')^{2}}{2t/m_{\rm{p}}}\Br]\exp \bl[-W(x'-x''+(\al-\beta)L,T) \br] \label{C1} . 
\eea
Note that the function $\exp [-W ]$ is obtained as
\bea
\exp [-W(x'-x'',T) ]=\bra{0_{\phi}} U^{\dg}(x',T)U(x'',T) \ket{0_{\phi}},
\eea
where $U(x,t)$ is the unitary operator defined by the equation (\ref{15}). The absolute value of $\exp [-W ]$ is smaller than $1$. Hence, the integrand 
\beq
\exp \Bl[-\fr{x'^{2}+x''^{2}}{2}\Br]\exp \bl[-W(x'-x''+(\al-\beta)L,T) \br]
\eeq
is exponentially small for $|x'| >\sq{2}$ and $|x''|>\sq{2}$. If the period $t/m_{\rm{p}}$ in the integrand of (\ref{C1}) is much larger than $1$, the main contribution of the integral (\ref{C1}) is obtained from the domain of the integration $|x'| \leq \sq{2}$ and $|x''| \leq \sq{2} $. Then, let us evaluate the function $W$ for $|x'-x''| \leq 1$. \\
~~~According to App.C, the function $W(x'-x''+(\al-\beta)L,T)$ is given by
\bea
W(x'-x''+(\al-\beta)L,T)&=&\fr{g^{2}}{\pi m^{2}}\int^{\infty}_{0} \fr{du}{(u^{2}+1)^{3/2}}(1-\cos[m(x-x'+(\al-\beta)L)u]) \nnm \\
&\times& (1-\cos[mT\sq{u^{2}+1}]) 
\eea
When $|\al-\beta|=0$, the function $W$ has the upper bound 
\bea
W(x'-x'',T)&=&\fr{g^{2}}{\pi m^{2}}\int^{\infty}_{0} \fr{du}{(u^{2}+1)^{3/2}}(1-\cos[m(x-x')u])(1-\cos[mT\sq{u^{2}+1}])\nnm \\
&\leq& \fr{2g^{2}}{\pi m^{2}}\int^{\infty}_{0} \fr{du}{(u^{2}+1)^{3/2}}(1-\cos[m(x-x')u]) \nnm \\
&=&\fr{2g^{2}}{\pi m^{2}}\Bl(1-m|x-x'|K_{1}(m|x-x'|)\Br), \label{D5}
\eea
where $K_{\nu}(z)$ is the modified Bessel function. The above upper bound (\ref{D5}) approaches to zero in the limit $m~(\geq m|x'-x''|) \rightarrow 0$. When $|\al-\beta|=2$, we fixes the parameters $gT$, $mT$ and $2mL$.  If $m$ is sufficiently smaller than the other parameters, then we can evaluate the function $W(x'-x''+(\al-\beta)L,T)$ for $|x'-x''| \leq 1$ as 
\bea
W(x'-x''+(\al-\beta)L,T) 
&=&\fr{g^{2}}{\pi m^{2}}\int^{\infty}_{0} \fr{du}{(u^{2}+1)^{3/2}}(1-\cos[m(x'-x''+(\al-\beta)L)u]) \nnm \\
&\times& (1-\cos[mT\sq{u^{2}+1}])  \nnm \\
&\approx&\fr{g^{2}}{\pi m^{2}}\int^{\infty}_{0} \fr{du}{(u^{2}+1)^{3/2}}(1-\cos[m|\al-\beta|Lu])\nnm \\
&\times& (1-\cos[mT\sq{u^{2}+1}])  \nnm \\
&=&W(2L,T).
\eea
Note that $W(0,T)=0$ and we can get the asymptotic formula of $p_{\al \beta}$ for  $t/m_{\rm{p}} \gg1$ and the sufficiently small $m$ as follows:
\bea
p_{\al \beta}(x,t)
&=&\fr{A^{2}}{2\pi^{3/2}t/m_{\rm{p}}}\int dx'dx''  \exp \Bl[-\fr{x'^{2}}{2}-i\fr{(x-\al L-x')^{2}}{2t/m_{\rm{p}}}\Br] \nnm \\
&\times& \exp \Bl[-\fr{x''^{2}}{2}+i\fr{(x-\beta L-x'')^{2}}{2t/m_{\rm{p}}}\Br]\exp \bl[-W(x'-x''+(\al-\beta)L,T) \br] \nnm \\
&\approx&\fr{A^{2}}{2\pi^{3/2}t/m_{\rm{p}}}\int^{\sq{2}}_{-\sq{2}} dx'\int^{\sq{2}}_{-\sq{2}}dx''  \exp \Bl[-\fr{x'^{2}}{2}-i\fr{(x-\al L-x')^{2}}{2t/m_{\rm{p}}}\Br] \nnm \\
&\times& \exp \Bl[-\fr{x''^{2}}{2}+i\fr{(x-\beta L-x'')^{2}}{2t/m_{\rm{p}}}\Br]\exp \bl[-W(x'-x''+(\al-\beta)L,T) \br] \nnm \\
&\approx&\fr{A^{2}}{2\pi^{3/2}t/m_{\rm{p}}}\exp \bl[-W(|\al-\beta|L,T) \br]  \nnm \\
&\times& 
\int^{\sq{2}}_{-\sq{2}} dx'\int^{\sq{2}}_{-\sq{2}}dx''  \exp \Bl[-\fr{x'^{2}}{2}-i\fr{(x-\al L-x')^{2}}{2t/m_{\rm{p}}}\Br] \exp \Bl[-\fr{x''^{2}}{2}+i\fr{(x-\beta L-x'')^{2}}{2t/m_{\rm{p}}}\Br] \nnm \\
&\approx&\fr{A^{2}}{2\pi^{3/2}t/m_{\rm{p}}}\exp \bl[-W(|\al-\beta|L,T) \br]  \nnm \\
&\times& 
\int dx'dx''  \exp \Bl[-\fr{x'^{2}}{2}-i\fr{(x-\al L-x')^{2}}{2t/m_{\rm{p}}}\Br] \exp \Bl[-\fr{x''^{2}}{2}+i\fr{(x-\beta L-x'')^{2}}{2t/m_{\rm{p}}}\Br], \label{D7}
\eea
where we have extended the domain of the integration from $[-\sq{2},\sq{2}]$ to $(-\infty,\infty)$ in the third approximation in the above equation (\ref{D7}). This is how we derive the equation (\ref{23}). 
{\flushleft{\section{Calculation of the integral in the equation (\ref{36})}}}
The remaining integral in the equation (\ref{36}) is calculated as 
\bea
\int_{-\infty}^{\infty}du \fr{1}{1+e^{-W}\cos u}e^{-\fr{u^{2}}{4L^{2}}}
&=&2L\int_{-\infty}^{\infty}du \fr{1}{1+e^{-W}\cos (2Lu)}e^{-u^{2}} \nnm \\
&=&2L \sum_{n=0}^{\infty} (-e^{-W})^{n} \int du (\cos( 2Lu) )^{n} e^{-u^{2}} \nnm \\
&=&2L \sum_{n=0}^{\infty} \Bl(-\fr{e^{-W}}{2}\Br)^{n} \sum^{n}_{k=0} \int^{\infty}_{-\infty} du~ {}_{n}C_{k} e^{2iL(n-2k)u} e^{-u^{2}} \nnm \\
&=&2L \sum_{n=0}^{\infty} \Bl(-\fr{e^{-W}}{2}\Br)^{n} \sum^{n}_{k=0}  {}_{n}C_{k}\sq{\pi} e^{-L^{2}(n-2k)^{2}}.  
\eea
For $2L \gg 1$, the above equation is approximated as
\beq
\approx 2L \sq{\pi} \sum_{n=0}^{\infty} \Bl(-\fr{e^{-W}}{2}\Br)^{2n}  {}_{2n}C_{n}=2L\sq{\pi} \fr{1}{\sq{1-e^{-2W}}}. 
\eeq
Therefore, in the limit $t/m_{\rm{p}} \rightarrow \infty$ and for $2L \gg 1$, we get the following formula of FI:
\bea
\mcl{F}_{ij}(t)&\rightarrow& \fr{A^{2}}{\sq{\pi}L}\Bl(\pa_{\tht_{i}}W\pa_{\tht_{j}}W\Br)\Bl(\int du\fr{\exp \bl[-\fr{u^{2}}{4L^{2}}\br]}{1+e^{-W}\cos u} +2L\sq{\pi}(e^{-W-L^{2}}-1)\Br) \nnm \\
&\approx& \Bl(\fr{1}{\sq{1-e^{-2W}}}-1\Br)\Bl(\pa_{\tht_{i}}W\pa_{\tht_{j}}W\Br),
\eea
where $A=(2+2e^{-L^{2}})^{-1/2}$.
\end{document}